\documentclass{jfmOld}
\usepackage{graphicx}
\usepackage{newtxtext}
\usepackage{newtxmath}
\usepackage{natbib}
\usepackage{hyperref}
\hypersetup{
    colorlinks = true,
    urlcolor   = blue,
    citecolor  = black,
}

\newcommand{\RomanNumeralCaps}[1]
\linenumbers

\usepackage{amsmath,amssymb}
\hypersetup{
	bookmarksnumbered,
	hidelinks
}

\graphicspath{{./Fig/},{./}}
\usepackage{placeins} % to use \FloatBarrier

\usepackage{cleveref}
\crefname{figure}{figure}{figures}
\crefname{equation}{}{}
\crefformat{section}{\S#2#1#3}
\crefmultiformat{section}{\S\S~#2#1#3}{ and~#2#1#3}{, #2#1#3}{, and~#2#1#3}
\title{Transient dispersion process of active particles}

\author{Weiquan Jiang\aff{1} 
 \and Guoqian Chen\aff{1}
 \corresp{\email{gqchen@pku.edu.cn}}}

\affiliation{\aff{1}Laboratory of Systems Ecology and Sustainability Science, College of Engineering, Peking University, Beijing 100871, China}

\begin{document}
\maketitle

\begin{abstract}
Active particles often swim in confined environments.
The transport mechanisms, especially the global one as reflected by the Taylor dispersion model, are of great practical interest to various applications.
For active dispersion process in confined flows,
previous analytical studies focused on the long-time asymptotic values of dispersion characteristics.
Only several numerical studies preliminarily investigated the temporal evolution.
Extending recent studies of Jiang \& Chen (\textit{J.\ Fluid Mech.}, vol.\ 877, 2019, pp.\ 1--34; \textit{J.\ Fluid Mech.}, vol.\ 899, 2020, A18), 
this work makes the first analytical attempt to investigate the transient process.
The temporal evolution of  the local distribution in the confined-section--orientation space, drift, dispersivity and skewness, is explored based on moments of distributions.
We introduce the biorthogonal expansion method for solutions because the classic integral transform method for passive transport problems is not applicable due to the self-propulsion effect.
Two types of boundary condition, the reflective condition and the Robin condition for wall accumulation, are imposed respectively.
% Results
A detailed study on spherical and ellipsoidal swimmers dispersing in a plane Poiseuille flow demonstrates the influences of the swimming, shear flow, wall accumulation and particle shape on the transient dispersion process after a point-source release.
The swimming-induced diffusion makes the local distribution reach its equilibrium state faster than that of passive particles. % can be shortened
Though the wall accumulation significantly affects the evolution of the local distribution and the drift,  the time scale to reach the Taylor regime is not obviously changed.
The shear-induced alignment of ellipsoidal particles can enlarge the dispersivity but has less influence on the drift and the skewness.
\end{abstract}

\begin{keywords}

\end{keywords}

%{\bf MSC Codes }  {\it(Optional)} Please enter your MSC Codes here

\section{Introduction}

Active particles, e.g.\ micromotors and motile micro-organisms, can harvest energy from the environment for self-propulsion, known as active Brownian motion {\citep{schweitzer_brownian_2003,romanczuk_active_2012}},
which is fundamentally different from pure translational Brownian motion of passive particles without swimming ability.
The transport mechanism of active particles is significant for various biological, environmental and chemical applications, such as algae cultivation {\citep{posten_design_2009,acien_photobioreactors_2017}}, remedies for harmful algal blooms \citep{durham_thin_2012,liu_effects_2012}, bioreactors for biofuels {\citep{chisti_biodiesel_2007,bees_mathematics_2014}} and cargo transport \citep{yasa_microalga-powered_2018,xiao_review_2019}.

Active particles often swim in confined environments, e.g.\ synthetic microswimmers in a micro-channel, or bacteria in the digestive tract.
Complicated interactions of active particles with physical boundaries play a key role in transport process and result in rich phenomena, such as wall scattering \citep{drescher_fluid_2011,kantsler_ciliary_2013}, circular trajectories \citep{berg_chemotaxis_1990,lauga_swimming_2006}, shear-induced trapping\citep{rusconi_bacterial_2014} and rheotaxis \citep{uspal_rheotaxis_2015,mathijssen_oscillatory_2019,brosseau_relating_2019}.
As one of the most well-known phenomena, micro-organisms such as spermatozoa and \textit{Escherichia coli} are found to accumulate near surfaces of confined domains \citep{rothschild_non-random_1963,berke_hydrodynamic_2008}.
To explain this accumulation feature, many theoretical models have been proposed, such as the far-field  and near-field hydrodynamic models {\citep{berke_hydrodynamic_2008,li_accumulation_2009,spagnolie_hydrodynamics_2012,sipos_hydrodynamic_2015}} and steric models considering inter-molecular forces like the van der Waals force \citep{li_amplified_2008,costanzo_transport_2012,contino_microalgae_2015,chilukuri_dispersion_2015}.
Besides, many researchers have imposed a mathematically simple Robin boundary condition (the third-type) for the probability density function (p.d.f.) of active particles in the position--orientation space {\citep{enculescu_active_2011,elgeti_wall_2013,ezhilan_transport_2015,jiang_dispersion_2019,alonso-matilla_transport_2019,berlyand_kinetic_2020,peng_upstream_2020}}.
Using this no-penetration condition for the probability flux at the boundaries, the wall accumulation phenomenon can be readily realized in numerical simulations \citep{ezhilan_transport_2015,bearon_trapping_2015,nili_population_2017}.

% Long time limit model
Because of the complex behaviours of active particles at the microscale, the transport characteristics at the macroscale have attracted practical attentions.
From a microscopic viewpoint, a high-dimensional Smoluchowski equation can be used to describe the transport process of swimmers in the position--orientation space, i.e.\ the phase space \cite{doi_brownian_1988}.
% analytical analyse
The computational expense of such a microscopic model is potentially huge, even for some special applications \citep{zeng_distribution_2018}.
To characterise the effective transport process only in the position space (at the macroscale),
simple macro-transport models have been proposed, by homogenizing the fast- and small-scale swimming processes.
The well-known model, P--K model, proposed by \citet{pedley_new_1990,pedley_hydrodynamic_1992},
uses a Fokker--Planck equation for the local p.d.f.\ of the swimming direction at each point in the position space and the active drift vector and the active translational dispersivity tensor are calculated based on the local p.d.f.\ associated with a correlation time coefficient.
Another known model, called the GTD model, uses the generalized Taylor dispersion theory {\citep{frankel_foundations_1989,hill_taylor_2002,hill_bioconvection_2005,bearon_spatial_2011}} to calculate the translational dispersivity tensor and gives some corrections for the P--K model for flows with strong shear rates.
Though these two models are widely applied in current studies \citep{croze_gyrotactic_2017,fung_bifurcation_2020}, they are only valid when the swimming scale is much smaller than the length scale of the confined environments \citep{bearon_spatial_2011}.

Furthermore, for active particles dispersing in confined flows such as the common Poiseuille flow and Couette flow, the one-dimensional macro-transport process in the longitudinal direction is of particular interest.
The pioneering work by {\citet{bees_dispersion_2010}} introduced the P--K model for high-concentrated suspensions of gyrotactic swimmers in a vertical downwelling pipe flow.
They derived the overall drift and dispersivity in the longitudinal direction based on the moment method by \citet{aris_dispersion_1956}.
Apart from the P--K model, \citet{bearon_biased_2012,croze_dispersion_2013,croze_gyrotactic_2017} applied the GTD model and gave more accurate results for the drift and dispersivity.
However, these result may fail when the separate-length-scale requirement of the P--K model and GTD model is missed, e.g.\ when the length scale of the confined section is comparable to that of the swimming, or the boundary effect cannot be neglected \citep{bearon_spatial_2011}.
Recently, \citep{jiang_dispersion_2019,jiang_dispersion_2020} constructed a more integrated average approach also based on the GTD theory and analytically derived the overall drift and dispersivity for very dilute suspensions.
This method gets rid of the separate-length-scale requirement of the P--K and GTD  models, and thus is adaptable for wide applications.
\citet{jiang_dispersion_2019} also considered the influence of wall accumulation on the dispersion process by introducing the Robin boundary condition for the p.d.f.\ \citep{elgeti_wall_2013,ezhilan_transport_2015,bearon_trapping_2015}.
\citet{peng_upstream_2020} analysed the accumulation effect on the upstream swimming (drift) based an orientation-moment expansion method \citep{saintillan_active_2013} and performed comparisons with the result by Brownian dynamics simulation.

%Transient and skewness 
The above studies on the dispersion process of active particles in confined flows mainly focused on the long-time asymptotic characteristics.
However, little analytical work has been done to address the transient process.
In fact, for active particles in unbounded position space,
e.g.\ particles swim freely in a two-dimensional confined thin film or a three-dimensional space, 
abundant studies have investigated the transient  diffusion process before the long-time diffusion limit \citep{howse_self-motile_2007,tenhagen_brownian_2011,tenhagen_brownian_2011a,zheng_non-gaussian_2013,sandoval_stochastic_2014}.
Three basis stages are found in the temporal evolution of the mean squared displacement ($\mathrm{MSD}$) of active particles with rotational diffusion motions: diffusive at the short time scale ($\mathrm{MSD} \sim t$, $t$ is the time),  ballistic during the intermediate time scales ($\mathrm{MSD} \sim t^2$), and finally again diffusive at the long time scale ($\mathrm{MSD} \sim t$) but with an enhanced dispersivity \citep{bechinger_active_2016}.
Because of the simplicity of the transport problem in a free space, the $\mathrm{MSD}$ of active particles can be theoretically derived, even for the case with a simple shear background flow \citep{tenhagen_brownian_2011a,sandoval_stochastic_2014}.
This anomalous diffusion (super-diffusion or sub-diffusion) process can be further analysed using non-Gaussian statistics, such as skewness and kurtosis \citep{tenhagen_brownian_2011,zheng_non-gaussian_2013}.
% porous media? alonso-matilla_transport_2019

However, for confined flows, the boundaries tremendously increase the complexity of the transport problem of active particles,  especially considering complicated swimming behaviours near boundaries such as the wall accumulation effect.
To the best of our knowledge, only some numerical studies, mainly using the Brownian dynamics simulation method, have addressed the transient active dispersion process in confined flows.
\citet{croze_dispersion_2013} investigated the dispersion of swimming algae in laminar and turbulent channel flows.
The temporal evolution of the drift, effective diffusivity and skewness was calculated statistically.
{\citet{chilukuri_dispersion_2015}} also calculate these dispersion characteristics using a simplified interaction model {\citep{chilukuri_impact_2014}} considering the influence of hydrodynamic interactions for wall accumulation.
\citet{apaza_ballistic_2016} focused on the hydrodynamic effects on the transient scale of the $\mathrm{MSD}$.
Other studies \citep{ghosh_self-propelled_2013,ao_active_2014,yariv_ratcheting_2014,sandoval_effective_2014,makhnovskii_effect_2019} have experimentally and numerically investigated the transient active dispersion process in a corrugated channel without background flow, considering the application of sorting particles by their self-propelled speeds.
Additionally, it is of considerable interest to systematically compare the transient active dispersion process with the classic  dispersion of passive particles \citep{lighthill_initial_1966,foister_diffusion_1980,latini_transient_2001,camassa_exact_2010,vedel_time-dependent_2014,taghizadeh_preasymptotic_2020}, to capture the differences of the approach to the Taylor dispersion regime \citep{chatwin_approach_1970,wu_approach_2014}.

This work is to make the first analytical attempt to investigate the transient dispersion process of active particles in confined flows.
Based on the GTD theory used in our previous studies \citep{jiang_dispersion_2019}, we introduce the biorthogonal expansion method \citep{brezinski_biorthogonality_1991} to calculate the temporal evolution of moments of the cross-sectional mean concentration distribution, 
and then the basic dispersion characteristics, such as the local distribution in the confined-section--orientation space, the drift, dispersivity and skewness, can be obtained and analysed in the initial transient stage.
The biorthogonal expansion method is often used to study the rheology of suspensions of particles \citep{strand_computation_1987,nambiar_stress_2019}.
As an extension of the classic integral transform method with orthogonal bases for passive transport problems, the biorthogonal expansion method can solve the difficulty caused by the effect of the self-propulsion for the active transport problems.
The auxiliary eigenvalue problem for the moments of distributions is solved by the Galerkin method with function series constructed for specific boundary conditions.
The typical reflective boundary condition \citep{bearon_spatial_2011,ezhilan_transport_2015} often used in numerical studies ideally assuming elastic collisions between the wall and the particles \citep{volpe_simulation_2014,bechinger_active_2016} is imposed.
To account for the wall accumulation phenomenon, we also consider the Robin boundary condition \citep{enculescu_active_2011,ezhilan_transport_2015}.
The rest of this paper is structured as follows.
For the active transport problem formulated in \cref{sec_formulation}, we introduce the definition of moments of the p.d.f.\ and the dispersion characteristics in \cref{sec_solution_moments}.
The corresponding governing equations are solved using the biorthogonal expansion method.
In \cref{sec_results}, a detailed study on the transient active dispersion process in a plane Poiseuille flow is demonstrated.
We focus on the influences of the swimming, shear flow, boundary effect (wall accumulation) and particle shape on the transient dispersion process.

\section{Formulation of transport problem}
\label{sec_formulation}

\subsection{Governing equations}

\begin{figure}
	\centering
	{\includegraphics{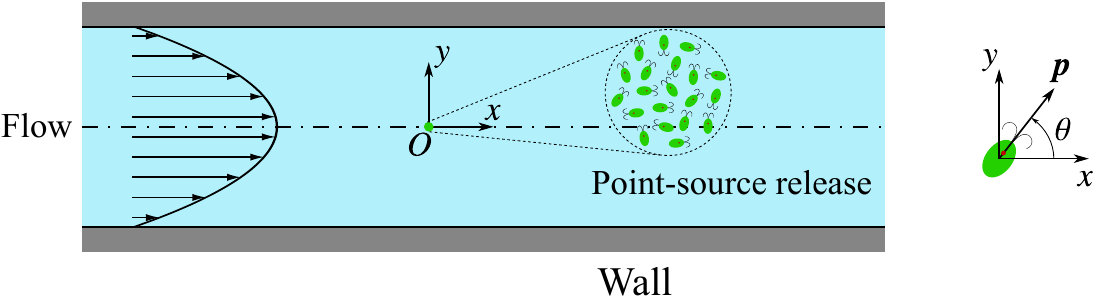}}
	\caption{ Sketch of a dilute suspension of active particles in a plane Poiseuille flow.
		\label{fig sketch}}
\end{figure}

As depicted in \cref{fig sketch}, we consider a very dilute suspension of active particles in a unidirectional flow between two planes.
The transport equation in the position--orientation space (phase space) \citep{doi_brownian_1988} can be adopted as
\begin{multline}
	\frac{\partial P}{\partial t} + \left[ \mathit{Pe}_f u (y) + \mathit{Pe}_s
	\cos \theta \right]  \frac{\partial P}{\partial x} + \mathit{Pe}_{\mathrm{s}}
	\sin \theta \frac{\partial P}{\partial y} + \frac{\partial}{\partial \theta} 
	[\Omega (y, \theta) P]
	\\
	= D_t \frac{\partial^2 P}{\partial x^2} + D_t
	\frac{\partial^2 P}{\partial y^2} + \frac{\partial^2 P}{\partial \theta^2},
	\label{eq probability conservation simple}
\end{multline}
where $t$ is the time, $x$ and $y$ are the position coordinates, $\theta$ is the angle between the swimming direction $\boldsymbol{p}$ of the particle and the longitudinal unit vector,
and $P(x,y,\theta,t)$ is the p.d.f..

Following \citet{jiang_dispersion_2019}, we introduce the following dimensionless variables and parameters (the superscript $\ast$ denotes dimensional variables) as
\begin{equation} \label{eq dimensionless variable}
	\left. \begin{gathered}
		t = t^{\ast} D^{\ast}_r, \quad 
		x = \frac{x^{\ast}}{W^{\ast}} - \mathit{Pe}_f t,  \quad 
		y = \frac{y^{\ast}}{W^{\ast}}, \quad 
		u = \frac{u^{\ast}}{u^{\ast}_m} -1 , \quad
		\\
		\Omega = \frac{\Omega^{\ast}}{D^{\ast}_r},\quad
		\mathit{Pe}_s = \frac{V_s^{\ast}}{D^{\ast}_r W^{\ast}}, \quad
		\mathit{Pe}_f = \frac{u^{\ast}_m}{D^{\ast}_r W^{\ast}}, \quad 
		D_t = \frac{D^{\ast}_t}{D^{\ast}_r (W^{\ast})^2},
	\end{gathered} \right\} 
\end{equation}
where $D^{\ast}_r$ is the rotational diffusion coefficient, $W^{\ast}$ is the channel width, $u(y)$ is the velocity profile, $u^{\ast}_m$ is the mean flow speed
\begin{equation}
	u^{\ast}_m \triangleq \frac{1}{W^{\ast}}  \int^{W^{\ast}}_0 u^{\ast}
	(y^{\ast}) \; {\mathrm{d}} y^{\ast}.
	\label{eq mean velocity}
\end{equation}
$\Omega$ is the angular velocity of $\theta$, $V_s^{\ast}$ is the swimming speed of the active particle,  $\mathit{Pe}_s$ is the corresponding swimming P{\'e}clet number,  $\mathit{Pe}_f$ is the flow
P{\'e}clet number, and $D_t$ is the ratio of the translational diffusivity to the rotational diffusivity.
We assume that the translational diffusivity is isotropic.
Note that the dimensionless velocity profile is the deviation from the mean flow speed because we have transformed the  frame of reference to that moving with the mean flow speed.

Due to the rotational and straining motion of the fluid, the rate of change of swimming direction for an ellipsoidal particle is given by Jeffery's equation {\citep{jeffery_motion_1922,leal_rheology_1972,pedley_hydrodynamic_1992,guazzelli_physical_2012}} as
\begin{equation}
	\Omega (y, \theta) = \frac{\mathit{Pe}_f}{2}  \frac{\mathrm{d} u}{\mathrm{d}
		y}  [-1 + \alpha_0 \cos (2 \theta)],
	\label{eq angular velocity}
\end{equation}
where $\alpha_0$ is the shape factor of the particle, with $\alpha_0 = 0$ for a spherical particle and $\alpha_0=1$ for an infinitely thin rod-like particle.

\subsection{Boundary conditions and initial condition}

For the solid boundaries, we consider two different types of condition.
First, the reflective condition assumes that collisions between particles and solid boundaries are perfectly
elastic \citep{bearon_spatial_2011,volpe_simulation_2014,jiang_dispersion_2019,jiang_dispersion_2020},
Thus, it requires that both the incident swimming probability flux and the incident transitional-diffusion probability flux through the walls are balanced by their corresponding reflective fluxes.
Namely,
\begin{equation}\label{eq_reflective_BC}
	\left. 
	\begin{aligned}
		P (x, y, \theta, t) & = P (x, y, - \theta, t), \quad \mathrm{at} \; y
		= 0, 1,
		\\
		\frac{\partial P}{\partial y} (x, y, \theta, t) & = - \frac{\partial
			P}{\partial y} (x, y, - \theta, t), \quad \mathrm{at} \; y = 0, 1,
	\end{aligned}
	\right\}
\end{equation}
ensuring the conservation of particles in the phase space.

Second,  we consider the equally typical Robin condition \citep{enculescu_active_2011,ezhilan_transport_2015,jiang_dispersion_2019} to account for the wall accumulation phenomenon of some kinds of active particles, e.g.\ sperm cells and \textit{E. coli} \citep{rothschild_non-random_1963,berke_hydrodynamic_2008}.
For each swimming direction, there is no penetration of the probability flux through the walls. Namely,
\begin{equation}
	D_t \frac{\mathrm{d} P}{\mathrm{d} y} = \mathit{Pe}_s \sin \theta P \quad
	\mathrm{at} \; y = 0, 1,
	\label{eq_Robin_BC}
\end{equation}
which is a third-type boundary condition.
To balance the incident swimming flux, the wall-normal transitional-diffusion flux must be negative, which leads to the accumulation of particles swimming towards a wall \citep{ezhilan_transport_2015}.
Note that this mechanism for wall accumulation does not consider the complicated hydrodynamic and steric interactions between particles and walls \citep{bechinger_active_2016,lauga_hydrodynamics_2009}.

In the orientation space, periodic boundary conditions are imposed
\begin{equation} \label{eq_periodic_BC}
	\left.
	\begin{aligned}
		P |_{\theta = -\upi} &=  P |_{\theta = \upi}, 
		\\
		\left. \frac{\partial
			P}{\partial \theta} \right|_{\theta = -\upi} &= \left. \frac{\partial P}{\partial
			\theta} \right|_{\theta = \upi} . 
	\end{aligned}
	\right\}
\end{equation}

For the initial condition, we consider particles released at the middle of the channel swimming in random directions, i.e.\
\begin{equation}
	P |_{t = 0} = \frac{1}{2 \upi} \delta\left( y - 0.5 \right),
	\label{eq_initial_condition}
\end{equation}
where $\delta(y)$ is the Dirac delta function.
There is no doubt that the initial condition will greatly affect the transient dispersion process of active particles but does not influence the long-time asymptotic behaviour.

\section{Solutions of transient dispersion characteristics}
\label{sec_solution_moments}

The dispersion process of active particles in the longitudinal direction is of particular interest because  the longitudinal scale is much larger than the transverse scale for a unidirectional confined flow.
Taking the longitudinal coordinate variable $x$ as the global space variable, and the confined section variables $y$ and $\theta$ as the local space variables, previous studies \citep{jiang_dispersion_2019,jiang_dispersion_2020} have applied the generalized Taylor dispersion theory \citep{brenner_general_1982a,brenner_macrotransport_1993} to analyse the long-time asymptotic values of dispersion characteristics, such as the local distribution, the drift and dispersivity.

In this work, we focus on the temporal evolution of these basic dispersion characteristics.
We first introduce the definition of the moments of p.d.f.\ and their governing equations.
Then we use the biorthogonal expansion method \citep{strand_computation_1987,brezinski_biorthogonality_1991} to solve the moments.
The auxiliary eigenvalue problem for the moments is solved by the Galerkin method with confined-section--orientation function series constructed for the reflective boundary condition and the Robin condition \citep{jiang_dispersion_2019} respectively.

\subsection{Moments and dispersion characteristics}

The dispersion characteristics are derived from the moments of the probability distribution of particles.
First, the moments of p.d.f.\ are conventionally defined as \citep{aris_dispersion_1956,brenner_macrotransport_1993}
\begin{equation}
	P_n (y, \theta, t) \triangleq \int^{\infty}_{- \infty} x^n P (x, y, \theta,
	t) \; \mathrm{d} x, \quad n = 0, 1, \ldots,
	\label{eq definition moment Pn}
\end{equation}
which are also called the local moments.
Note that the zeroth-order moment, $P_0$,  is the marginal p.d.f.\ in the cross-section ($\{(y, \theta)\}$) of the phase space, and thus  can be viewed as the local distribution of active particles \citep{ezhilan_transport_2015,jiang_dispersion_2019}.

Second, we introduce the global moments, i.e.\ the moments of the cross-sectional mean concentration distribution $\bar{P}$,
\begin{equation}
	M_n (t) \triangleq \int^{\infty}_{- \infty} x^n \bar{P} 
	\; \mathrm{d} x = \bar{P_n}  \quad n = 0, 1, \ldots,
	\label{eq definition moment Mn}
\end{equation}
and 
\begin{equation}
	\bar{P}(x,t) \triangleq \int^1_0 \int^{\upi}_{-\upi} P (x, y, \theta, t) \;
	\mathrm{d} \theta \mathrm{d} y.
	\label{eq definition operation integration}
\end{equation} 
We use the bar to denote the integration over the cross-section ($\{(y, \theta)\}$).
Due to the conservation of particles, we have $M_0=1$.

The basic dispersion characteristics, i.e.\ the drift $U_d$ and dispersivity $D_T$, are related to the first- and second-order global moments,
\begin{align}
	U_d (t) & \triangleq  \frac{\mathrm{d} \mu_x}{\mathrm{d} t} =
	\frac{\mathrm{d} M_1}{\mathrm{d} t},  
	\label{eq_def_drift}
	\\
	D_T (t) & \triangleq  \frac{1}{2}  \frac{\mathrm{d} \sigma^2}{\mathrm{d} t}
	= \frac{1}{2}  \frac{\mathrm{d} M_2}{\mathrm{d}
		t} - M_1  \frac{\mathrm{d} M_1}{\mathrm{d} t},  
	\label{eq_def_dispersivity}
\end{align}
where $\mu_x$ and $\sigma$ are the expected value (mean displacement) and the standard deviation ($\mathrm{MSD}$) respectively:
\begin{align}
	\mu_x & \triangleq \frac{M_1}{M_0} = M_1, 
	\\ 
	\sigma^2 & \triangleq \frac{M_2}{M_0} - \frac{M^2_1}{M^2_0} = M_2 - M_1^2 .
\end{align}
Their long-time asymptotic values correspond to the coefficients used in the famous Taylor dispersion model \citep{taylor_dispersion_1953,taylor_dispersion_1954}.
Thus, their temporal evolution can outline the longitudinal transport process in the transient stage before the Taylor dispersion regime \citep{gill_note_1967,gill_exact_1970,chatwin_approach_1970,latini_transient_2001,wu_approach_2014}.

Apart from the above basic dispersion characteristics, one can also introduce the skewness of p.d.f., to capture the asymmetry of distribution, especially in the initial stage after particle release \citep{chatwin_approach_1970,wang_basic_2017,jiang_solute_2019}.
The skewness $\gamma_1$ is defined by the third-order cumulant $\kappa_3$ of the distribution as
\begin{equation} \label{eq_def_skewness}
	\gamma_1 \triangleq \frac{\kappa_3}{\sigma^3},
\end{equation}
where
\begin{equation}
	\kappa_3 \triangleq \frac{M_3}{M_0} - 3 \frac{M_2 M_1}{M_0^2} + 2 \frac{M_1^3}{M_0^3} 
	= M_3 -3 M_2 M_1 +2 M_1^3
	.
\end{equation}

\subsection{Solutions of moments: biorthogonal expansion}

\subsubsection{Governing equation of moments}

To obtain the transient dispersion characteristics, we solve the moments first.
According to the definition of moments \cref{eq definition moment Pn} and the governing equation of the p.d.f\  \cref{eq probability conservation simple}, with the assumption that the p.d.f.\ decays exponentially as $| x |\rightarrow \infty$ {\citep{aris_dispersion_1956}}, 
we have
\begin{equation}
	\frac{\partial P_n}{\partial t} +\mathcal{L}P_n = n (n - 1)  D_t P_{n - 2} + n \left[ \mathit{Pe}_{{f}} u
	(y) + \mathit{Pe}_{{s}} \cos \theta \right] P_{n - 1}, \quad n = 0, 1, \ldots,
	\label{eq_moment_governing_Pn}
\end{equation}
where $P_{- 1} = P_{- 2} = 0$ and 
\begin{equation}
	\mathcal{L} (\cdot) \triangleq \mathit{Pe}_{{s}} \sin \theta
	\frac{\partial}{\partial y} (\cdot) + \frac{\partial}{\partial \theta}
	\left[ \Omega (y, \theta) (\cdot) - \frac{\partial}{\partial \theta} (\cdot)
	\right] - D_t \frac{\partial^2}{\partial y^2}
	(\cdot)
	\label{eq_def_L_operator}
\end{equation}
is an operator corresponding to the transport equation in the cross-section.

The boundary conditions of $P_n$ ($n = 0, 1, \ldots$) are in the same form as those of $P$.
Namely, for the reflective condition \cref{eq_reflective_BC},
\begin{equation} \label{eq_Pn_reflective_BC}
	\left.
	\begin{aligned}
		P_n (y, \theta, t) & =  P_n (y, - \theta, t), \quad \mathrm{at} \; y
		= 0, 1,
		\\
		\frac{\partial P_n}{\partial y} (y, \theta, t) & =  - \frac{\partial
			P_n}{\partial y} (y,  - \theta, t), \quad \mathrm{at} \; y = 0, 1.
	\end{aligned}
	\right\}
\end{equation}
For the Robin condition \cref{eq_Robin_BC},
\begin{equation}
	D_t \frac{\mathrm{d} P_n}{\mathrm{d} y} = \mathit{Pe}_s \sin \theta P_n \quad
	\mathrm{at} \; y = 0, 1.
	\label{eq_Pn_Robin_BC}
\end{equation}
In the orientation space,
\begin{equation}\label{eq_Pn_periodic_BC}
	\left.
	\begin{aligned}
		P_n |_{\theta = -\upi} &=  P_n |_{\theta = \upi}, 
		\\
		\left. \frac{\partial
			P_n}{\partial \theta} \right|_{\theta = -\upi} &= \left. \frac{\partial P_n}{\partial
			\theta} \right|_{\theta = \upi} . 
	\end{aligned}
	\right\}
\end{equation}
The initial conditions are
\begin{align}
	P_0 |_{t = 0} & =  \frac{1}{2 \upi} \delta (y - 0.5), 
	\\
	P_n |_{t = 0} & =  0, \quad n = 1, 2, \ldots.
\end{align}

We can also obtain the governing equation for the global moments.
Note that according to the integration by parts formula, we have 
\begin{equation*}
	\overline{\mathcal{L}P_n} = 0, \quad n = 0, 1, \ldots,
\end{equation*}
under both the reflective condition \labelcref{eq_Pn_reflective_BC} and the Robin condition \labelcref{eq_Robin_BC}.
Therefore,
\begin{equation}\label{eq_moment_governing_Mn}
	\frac{\mathrm{d} M_n}{\mathrm{d} t} = n (n - 1) D_t M_{n - 2} + n
	\overline{\left( \mathit{Pe}_f u + \mathit{Pe}_s \cos \theta \right) P_{n -
			1}}, \quad n = 1,2 \ldots.
\end{equation}
In particular, 
\begin{equation}\label{eq_moment_governing_drift}
	U_d = \overline{\left( \mathit{Pe}_f u + \mathit{Pe}_s \cos \theta \right) P_0},
\end{equation}
namely, the local-distribution-weighted average of the longitudinal velocity component.

Note that the form of moment equation \cref{eq_moment_governing_Pn} is similar to that of the case of passive particles.
Previous studies used the method of separation of variables or the integral transform method \citep{barton_method_1983,jiang_solute_2019} to derive a series expansion for the solutions.
An auxiliary Sturm--Liouville problem was solved first to obtain the function basis for the expansion.

However, for the present case of active particles, the local operator $\mathcal{L}$ \cref{eq_def_L_operator} associated with the boundary conditions can be non-self-adjoint.
The method of separation of variables and the classic integral transform method are not feasible.
Instead, we use the biorthogonal expansion method (an extension of the integral transform method) \citep{strand_computation_1987,brezinski_biorthogonality_1991,nambiar_stress_2019} to obtain series expansions for the local moments and the Galerkin method to solve the associated eigenvalue problem.
Two different function bases are used in the Galerkin method for the reflective condition and the Robin condition respectively.

\subsubsection{Biorthogonal expansion}

The auxiliary eigenvalue problem for the moment equation \cref{eq_moment_governing_Pn} is
\begin{equation}
	\mathcal{L} f_i = \lambda_i f_i,
	\label{eq_eigenvalue_problem}
\end{equation}
where $\lambda_i$ is the eigenvalue ($i=1,2,\ldots,$) and $f_i$ is the associated eigenfunction satisfying all the boundary conditions of $P_n$.
For $\lambda_1=0$, $f_1$ corresponds to the long-time asymptotic solution of $P_0$, which was discussed in our previous paper \citep{jiang_dispersion_2019}.
It is difficult to find the explicit expression of the solution of the associated eigenfunction $f_i$, due to the complexity of $\mathcal{L}$ \cref{eq_def_L_operator}.
We use the Galerkin method to approximately solve  $\lambda_i$ and $f_i$.

Suppose we have found a basis with functions satisfying the required boundary conditions.
Detailed expressions of the bases for the reflective condition and the Robin condition are later shown in \cref{sec_basis_functions}.
Now with such a basis, denoted by $\{g_i\}_{i=1}^{\infty}$, we can expand the eigenfunction $f_i$ as
\begin{equation}
	f_i = \sum_{j = 1}^{\infty} \phi_{i j} g_j,
	\label{eq_f_i_expansion}
\end{equation}
where $ \phi_{i j}$ is the coefficient of the expansion.
For the local operator $\mathcal{L}$, we can also express the corresponding bilinear form $A(\cdot, \cdot)$ with the basis.
The elements of the corresponding matrix are
\begin{equation}
	\mathsfbi{A}_{i j} = A(g_i, g_j) = \langle g_i, \mathcal{L} g_j \rangle, \quad i=1,2,\ldots, \;  j=1,2,\ldots,
\end{equation}
where $\langle \cdot, \cdot \rangle$ denotes the associated inner product.
In matrix form, the weak formulation of the auxiliary eigenvalue problem \cref{eq_eigenvalue_problem} can be written as
\begin{equation}\label{eq_eigenvalue_problem_matrix}
	\mathsfbi{A} \boldsymbol{\phi}_i= \lambda_i  \boldsymbol{\phi}_i,
\end{equation}
where ${\boldsymbol{\phi}}_i =\begin{pmatrix} \phi_{i 1}, & \phi_{i 2}, & \cdots \end{pmatrix} ^{\mathrm{T}}$ is the vector of the coefficients of $f_i$.
Truncating the series \cref{eq_f_i_expansion}  to some degree $N$ gives
a Galerkin solution for the eigenfunction $f_i$.

Note that $\lambda_i$ is the eigenvalue of the matrix $\mathsfbi{A}$ and ${\boldsymbol{\phi}}_i$ is  the corresponding eigenvector.
Therefore, solving the eigenvalue problem of $\mathsfbi{A}$ can give asymptotic solutions of the eigenvalues and eigenfunctions of \cref{eq_eigenvalue_problem}.
In fact, the set of solutions $\{f_i\}_{i=1}^{N}$ can also form a basis for the function space satisfying the boundary conditions of $P_n$.
The corresponding transformation matrix from $\boldsymbol{g} \triangleq \begin{pmatrix} g_1, & g_2, & \cdots ,& g_N \end{pmatrix} ^{\mathrm{T}}$  to $\boldsymbol{f} \triangleq \begin{pmatrix} f_1, & f_2, & \cdots ,& f_N \end{pmatrix} ^{\mathrm{T}}$ is
\begin{equation}
	\mathsfbi{B} = \begin{pmatrix}
		{\boldsymbol{\phi}}_1, & {\boldsymbol{\phi}}_2, &
		\cdots, & {\boldsymbol{\phi}}_N
	\end{pmatrix},
\end{equation}
and then ${\boldsymbol{f}}^{\mathrm{T}} ={\boldsymbol{g}}^{\mathrm{T}} \mathsfbi{B}$.

With the eigenvalue $\lambda_i$ and eigenfunction $f_i$ solved, one can easily follow the work of \citet{barton_method_1983} and expand the local moments as
\begin{equation}\label{eq_Pn_series_expanion}
	P_n (y, \theta, t) = \sum_{i = 1}^{\infty} p_{n i} (t) \mathrm{e}^{\lambda_i t} f_i (y, \theta), \quad n=0,1,\ldots,
\end{equation}
where $p_{n i} (t)$ is the expansion coefficients.
Using  the method of separation of variables (or the integral transform),
\citet{barton_method_1983} derived the general expressions for the expansion coefficients $p_{n i} (t)$ (for $n$ up to three) with the elements of the bilinear form defined using the velocity profile and the initial condition.
See \S3 in his paper.
However, for the present case, the local operator $\mathcal{L}$ \cref{eq_def_L_operator} associated with the boundary conditions can be non-self-adjoint due to the swimming ($\mathit{Pe}_s \cos \theta$) and the angular velocity of active particles.
In fact, the eigenvalue of the matrix $\mathsfbi{A}$ of the local operator can be non-symmetric, resulting in complex eigenvalues and eigenvectors.
Thus the set of functions $\{f_i\}_{i=1}^{N}$ is not orthogonal, i.e.\ the inner product
\begin{equation*}
	\langle f_i, f_j\rangle \neq 0, \quad \text{for} \; i \neq j.
\end{equation*}
The orthogonality relation fails when applying the integral transform method to obtain $p_{n i} (t)$.

Instead of using the orthogonality relation, one can find another set of functions which bears a so-called biorthogonality relation with $\{f_i\}_{i=1}^{N}$.
According to the biorthogonal expansion method \citep{strand_computation_1987,brezinski_biorthogonality_1991}, the dual basis functions $f^{\star}_i$ (a superscript $\star$ denotes the dual counterpart) are the eigenfunctions of the adjoint operator of $\mathcal{L}$ (denoted $\mathcal{L}^{\star}$). 
After normalization, the biorthogonality relation is
\begin{equation}
	\langle f^{\star}_i, f_j\rangle = \delta_{i j},
\end{equation}
where $\delta$ is the Kronecker delta.
We can also use the Galerkin method to solve $f^{\star}_i$.
Let $\mathsfbi{A}^{\star}$ denote the corresponding matrix of $\mathcal{L}^{\star}$ is the transpose of $\mathsfbi{A}$,
then we have 
\begin{equation}
	\mathsfbi{A}^{\star} {\boldsymbol{\phi}}_i^{\star} = \lambda_i {\boldsymbol{\phi}}_i^{\star},
\end{equation}
where $ {\boldsymbol{\phi}}_i^{\star}$ is the coefficient vector of the solution for $f_i^{\star}$.
Performing the series expansion using the same basis as\cref{eq_f_i_expansion}, we have
\begin{equation}
	f^{\star}_i = \sum_{i = 1}^N \phi^{\star}_{i j} g_j
\end{equation}
and ${\boldsymbol{\phi}}_i^{\star} =\begin{pmatrix}
	\phi^{\star}_{i 1}, & \phi^{\star}_{i 2}, & \cdots, & \phi^{\star}_{i N}
\end{pmatrix}^{\mathrm{T}}$.
Note that the eigenvalues of $\mathsfbi{A}^{\star}$ are the same as those of $\mathsfbi{A}$ \citep{strand_computation_1987}.
In fact,  $\{f^{\star}_i\}_{i=1}^{N}$, the dual set of solutions $\{f_i\}_{i=1}^{N}$, can also form a basis. 
Let ${\boldsymbol{f}}^{\star} \triangleq \begin{pmatrix} f^{\star}_1, & f^{\star}_2, & \cdots ,& f^{\star}_N \end{pmatrix} ^{\mathrm{T}}$.
The corresponding transformation matrix from ${\boldsymbol{g}}$ is
\begin{equation}
	\mathsfbi{B}^{\star} = 
	\begin{pmatrix}
		{\boldsymbol{\phi}}_1^{\star}, & {\boldsymbol{\phi}}_2^{\star}, &
		\cdots, & {\boldsymbol{\phi}}_N^{\star}
	\end{pmatrix},
\end{equation}
and thus ${\boldsymbol{f}}^{\star \mathrm{T}} ={\boldsymbol{g}}^{\mathrm{T}} \mathsfbi{B}^{\star}$.
After normalization and using the biorthogonality relation, we have
\begin{equation}
	{\boldsymbol{f}}^{\mathrm{T}} {\boldsymbol{f}}^{\star}
	= \mathsfbi{B} {\boldsymbol{g}}^{\mathrm{T}}
	{\boldsymbol{g}} \mathsfbi{B}^{\star} = \mathsfbi{B} \mathsfbi{I} \mathsfbi{B}^{\star} = \mathsfbi{B}
	\mathsfbi{B}^{\star} = \mathsfbi{I},
\end{equation}
where $\mathsfbi{I}$ is the identity matrix.
Namely, $\mathsfbi{B}^{\star}$, comprised of the duel eigenvectors ${\boldsymbol{\phi}}_i^{\star}$, is the inverse of $ \mathsfbi{B}$.

With the biorthogonal family $\{\boldsymbol{f}, \boldsymbol{f}^{\star}\}$, one can continue to use the expressions obtained by \citet{barton_method_1983} for the expansion coefficients of moments in \cref{eq_Pn_series_expanion}, just by replacing the orthogonality relation with the biorthogonal one.
Namely, the matrix of the bilinear form $w_u(\cdot, \cdot)$ defined by the velocity profile is changed to
\begin{equation}
	w_u (f_i^{\star}, f_j) = \langle f_i^{\star} (y, \theta), u f_j\rangle.
\end{equation}
The initial values of $p_{n i}$ are 
\begin{align}
	p_{0 i} (0) &= \langle f^{\star}_i, \frac{1}{2 \upi} \delta (y - 0.5)
	\rangle, \quad i = 1, 2, \ldots,
	\\
	p_{n i} (0) &= 0, \quad i = 1, 2, \ldots, \; n = 1, 2, \ldots,
\end{align} 

Once we obtain the time-dependent solutions of the moments, the corresponding dispersion characteristics can be calculated according to their definitions without difficulties.
The last problem is to find the basis functions satisfying the boundary conditions of moments.

\subsubsection{Basis functions}
\label{sec_basis_functions}

First, we discuss the case with the reflective condition \cref{eq_Pn_reflective_BC}.
A reflective basis can be constructed  using the method of separation of variables for the  Laplace operator for the  transport equation of active particles in a tube \citep{jiang_dispersion_2020}.
Similarly, for the two-dimensional channel, a much simpler reflective basis can also be found for the Laplace operator, which is self-adjoint with respect to the reflective condition.
The basis is comprised of 
\begin{equation} \label{eq_reflective_basis}
	\frac{1}{\sqrt{2 \upi}}, \quad \frac{1}{\sqrt{\upi}} \cos (n \upi y), \quad
	\sqrt{\frac{2}{\upi}} \cos (n \upi y) \cos (m \theta), \quad
	\sqrt{\frac{2}{\upi}} \sin (n \upi y) \sin (m \theta),
\end{equation}
where $n=1,2,\ldots$ and $m=1,2,\ldots$.
A detailed derivation can be found in the paper of \citet{wang_vertical_2020}.
The corresponding inner product is just the $L^2$ inner product, i.e.\
\begin{equation}
	\langle f, g \rangle \triangleq \int^1_0 \int^{\upi}_{- \upi} f (y, \theta) g
	(y, \theta) \; \mathrm{d} \theta \mathrm{d} y,
\end{equation}
where $f$ and $g$ are functions that belong to the reflective basis.

Second, for the Robin condition \cref{eq_Pn_Robin_BC}, the construction of a basis in much more complicated, due to the swimming term with the coefficient $ \mathit{Pe}_s \sin \theta$.
Following \citet{jiang_dispersion_2019}, a decomposition form for the moments is applied before using the method of separation of variables:
\begin{equation}
	P_n (y, \theta) = P_a (y, \theta) G_n (y, \theta), \quad n = 0, 1, \ldots,
\end{equation}
where 
\begin{equation}
	P_a (y, \theta) = \exp \left[ \frac{\mathit{Pe}_s}{D_t} \left( y - \frac{1}{2} \right) \sin \theta  \right]
\end{equation}
satisfies the Robin condition \cref{eq_Pn_Robin_BC}, and $G_n (y, \theta)$ is modified moments satisfying a governing equation similar to \cref{eq_moment_governing_Pn}.
A detailed discussion can be found in \S 5 in that paper.
Note that the solid boundary condition is then changed from the Robin condition \cref{eq_Pn_Robin_BC} to a  Neumann condition (the second-type boundary condition),
\begin{equation}
	\left. \frac{\partial G_n}{\partial y} \right|_{y = 0, 1} = 0, \quad n = 0,
	1, \ldots.
\end{equation}
In the orientation space, $G_n$ satisfies the same periodic condition as \cref{eq_Pn_periodic_BC}.
Using the method of separation of variables of the Laplace operator for $G_n$, the basis for the Robin condition can be constructed as
\begin{equation} \label{eq_Robin_basis}
	\frac{P_a}{\sqrt{2 \upi}}, \quad \frac{P_a}{\sqrt{\upi}} \cos (n \upi y), \quad
	\sqrt{\frac{2}{\upi}} P_a \cos (n \upi y) \cos (m \theta), \quad
	\sqrt{\frac{2}{\upi}} P_a \cos (n \upi y) \sin (m \theta) .
\end{equation}
The corresponding inner product is defined with a weight function as
\begin{equation}
	\langle f, g \rangle \triangleq \int^1_0 \int^{\upi}_{- \upi} \frac{1}{P_a^2 
		(y, \theta)^{}} f (y, \theta) g (y, \theta) \; \mathrm{d}
	\theta \mathrm{d} y,
\end{equation}
where $f$ and $g$ are functions that belong to the Robin basis.

In the calculation of the Galerkin method, for both the reflective basis \cref{eq_reflective_basis} and the Robin basis \cref{eq_Robin_basis}, we collect terms with $n\leqslant20$ and $m\leqslant10$ to solve the eigenvalue problem \cref{eq_eigenvalue_problem_matrix}.
The total numbers of basis functions are $431$ and $441$ respectively.
For the biorthogonal expansion of moments \cref{eq_Pn_series_expanion}, we truncate the series with the upper bound of summation equal to $40$ to reduce the truncation error of the series expansion in the initial stage of the transport process.
The terms are sorted by the real part of the complex eigenvalue because higher-order terms decay much more rapidly.
The result by the biorthogonal expansion is verified with the numerical result by Brownian dynamics simulation, as shown in \cref{sec_simulations}.
We solve the first fourth moments.
The related dispersion characteristics, i.e.\ the drift $U_d$ \cref{eq_def_drift}, dispersivity $D_T$ \cref{eq_def_dispersivity} and skewness $\gamma_1$ \cref{eq_def_skewness}, are obtained accordingly.

%\subsection{Mean concentration distribution}

\section{Results}
\label{sec_results}

To compare the transient dispersion process of active particles with that of passive ones, we consider the case 
of active particles dispersing in a common plane Poiseuille flow.
The dimensionless velocity profile is $u(y) = 6 y (1-y) -1$.
Previous studies \citep{jiang_dispersion_2019,wang_vertical_2020} already discussed the long-time asymptotic values of dispersion characteristics, e.g.\ the local distribution, drift and dispersivity.
Here, we analyse the temporal evolution of these characteristics, as well as the skewness.
We focus on the influences of the swimming, shear flow, boundary effect (wall accumulation) and particle shape on the transient dispersion process.
In the following studied cases, we fix the translation diffusion coefficient $D_t=\frac{1}{6}$ based on the data of previous studies \citep{ezhilan_transport_2015,nili_population_2017,jiang_dispersion_2019}.
We mainly discuss spherical particles ($\alpha_0=0$) for simplicity, while the shear-induced alignment of  ellipsoidal particles is considered in \cref{sec_results_shape}
Additionally, a comparison with the numerical result by the Brownian dynamics simulation is presented in \cref{sec_simulations}.

\subsection{Influence of swimming}
\label{sec_results_swimming}

To analyse the swimming effect on the transient dispersion process, we consider spherical particles with different swimming ability.
Namely, the swimming P{\'e}clet numbers $\mathit{Pe}_s$ are different and $\mathit{Pe}_s=0$ corresponds to the case of passive particles.
To highlight the influence of swimming, there is no background shear flow (with the flow P{\'e}clet number $\mathit{Pe}_f =0$) and only the reflective boundary condition \cref{eq_Pn_reflective_BC} are considered.

\subsubsection{Local distribution: zeroth-order moment}

\begin{figure}
	\centering
	{\includegraphics{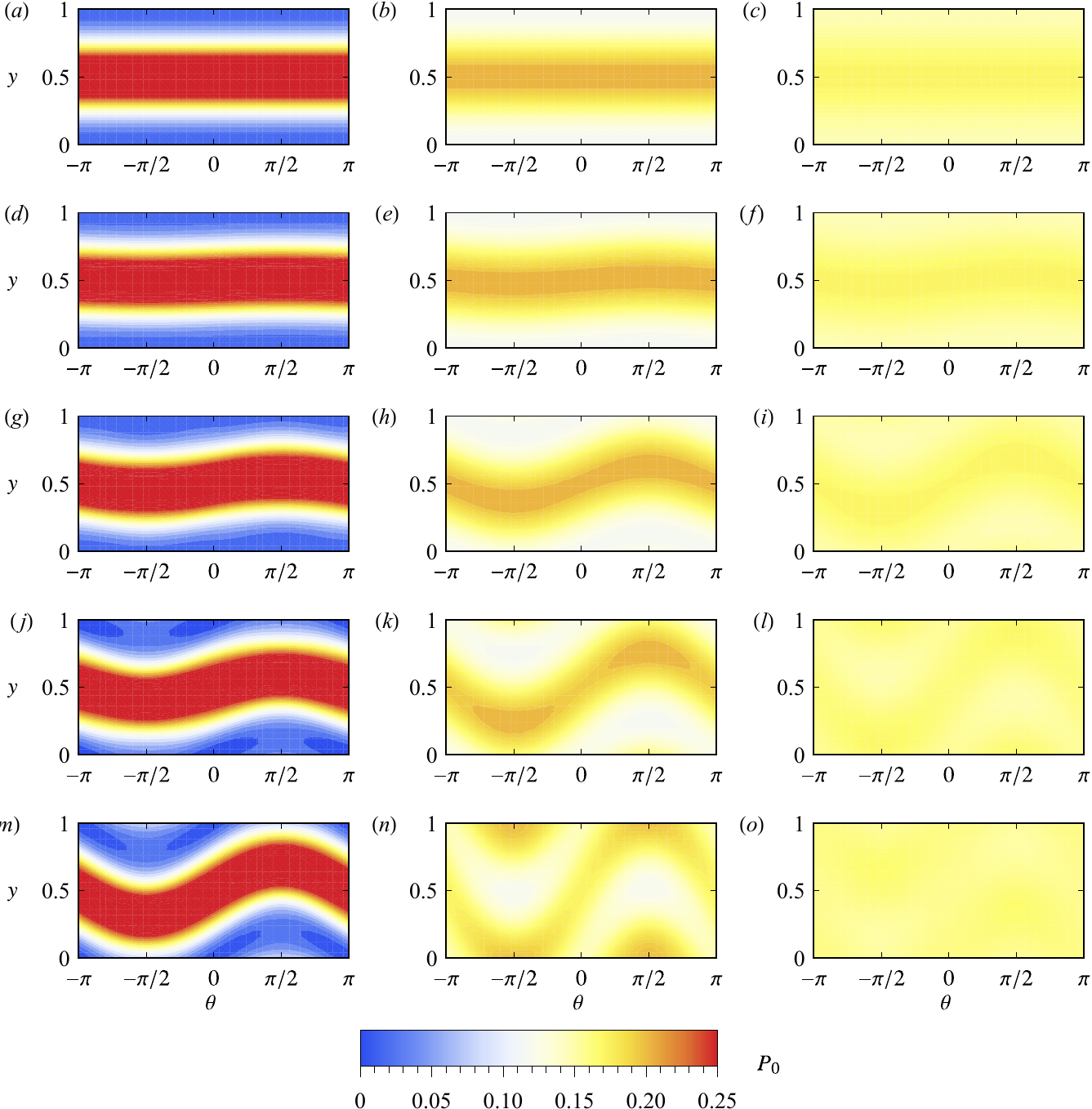}}
	\caption{Density plot of transient local distributions $P_0(y,\theta,t)$ of spherical particles with different swimming ability under the reflective condition.
		The swimming P{\'e}clet number: 
		(\textit{a}--\textit{c}) $\mathit{Pe}_s =0$; (\textit{d}--\textit{f}) $\mathit{Pe}_s =0.1$; (\textit{g}--\textit{i}) $\mathit{Pe}_s =0.5$; (\textit{j}--\textit{l}) $\mathit{Pe}_s =1$; (\textit{m}--\textit{o}) $\mathit{Pe}_s =2$.
		Sample times: (\textit{a},\textit{d},\textit{g},\textit{j},\textit{m}) $t=0.1$; (\textit{b},\textit{e},\textit{h},\textit{k},\textit{n}) $t=0.3$;
		(\textit{c},\textit{f},\textit{i},\textit{l},\textit{o}) $t=0.5$.
		In all cases, $\mathit{Pe}_f=0$.
		\label{fig_2D-P0-phi-theta}
	}
\end{figure}

%  initial stage
As shown in \cref{fig_2D-P0-phi-theta}, to depict the temporal evolution of the local distribution, $P_0$ at 3 small sample times ($t\in\{0.1, 0.3, 0.5\}$) are plotted.
As expected, the local transport process of active particles is greatly different from that of passive particles.
Without swimming, passive particles perform pure translational Brownian motions, while
the rotational diffusion of the ``swimming'' direction takes no effect due to the uniform initial distribution.
As shown in \cref{fig_2D-P0-phi-theta}(\textit{a}--\textit{c}), the distribution for $\theta$ is uniform, while in the transverse direction, the distribution become more and more uniform as particles spread out gradually.

% transient stage
For the active particles, the local transport process is a combination of the swimming motion and translational diffusion.
As shown in \cref{fig_2D-P0-phi-theta}(\textit{d}--\textit{o}), the swimming of particles leads to a sinusoidal variation of the distribution in the $O y \theta$ plane. 
After released in random directions, particles swim towards walls, resulting in a depletion of distribution in the middle of the channel during the transient transport process, as shown in \cref{fig_2D-P0-phi-theta}(\textit{k},\textit{n}) for particles with large swimming speeds.
Meanwhile, the rotational diffusion of the swimming direction  leads to the swimming-induced diffusion process and makes the distribution of $\theta$ uniform again.
Moreover, in \cref{fig_2D-P0-phi-theta}(\textit{m},\textit{n}), the reflection of the swimming probability flux at channel walls is  observed, as a result of the elastic collisions described by the reflective boundary condition \cref{eq_reflective_BC}.
Particles swim through the wall (e.g.\ $-\upi<\theta<0$ at $y=0$) is reflected back to the bulk in the reversed direction ($-\theta$).

% large times
Both the local distributions of active and passive particles become uniform in the whole local space as time increases.
Even when $t=0.5$, as shown in \cref{fig_2D-P0-phi-theta}(\textit{c},\textit{f},\textit{i},\textit{l},\textit{o}), the distributions are very uniform.
The results at larger times, not shown here, nearly have no difference between each other.
In fact, in the long-time limit, the local distribution of spherical particles is exactly uniform \citep{jiang_dispersion_2019}.
Obviously, the distribution of particles with stronger swimming ability will reach the uniform equilibrium faster, due to the swimming-induced diffusion effect.

\begin{figure}
	\centering
	{\includegraphics{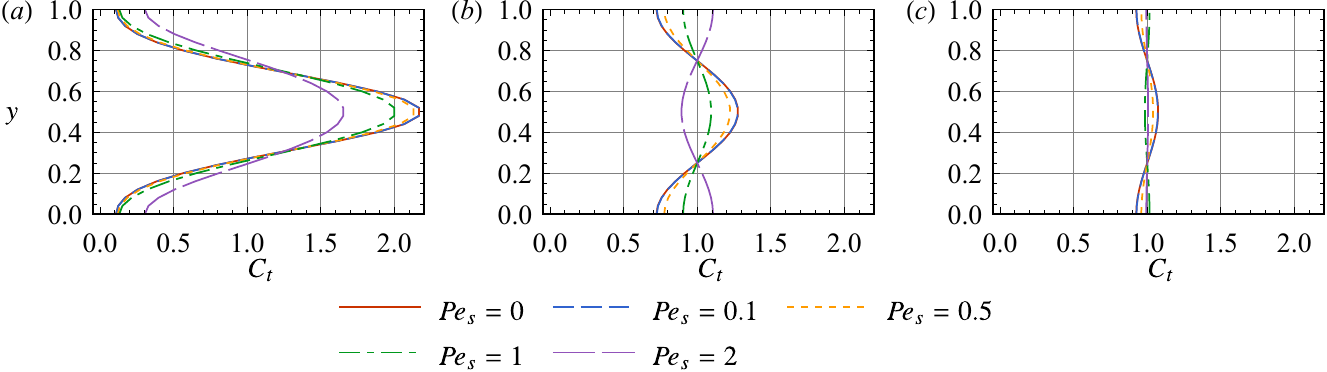}}
	\caption{Transverse distributions $C_t(y,t)$ of spherical particles with different swimming ability under the reflective condition.
		Sample times: (\textit{a}) $t=0.1$; (\textit{b}) $t=0.3$;
		(\textit{c}) $t=0.5$.
		In all cases, $\mathit{Pe}_f=0$.
		\label{fig_P0-z}
	}
\end{figure}

The swimming-induced diffusion effect on the local transport process can be demonstrated more clearly with the transverse distribution, defined as
\begin{equation} \label{eq_def_transverse_distribution}
	C_t (y, t) \triangleq \int^{\upi}_{- \upi} P_0 (y, \theta, t) \; \mathrm{d}
	\theta .
\end{equation}
As shown in \cref{fig_P0-z}, the larger the $\mathit{Pe}_s$, the smaller the concentration gradient.
At $t=0.5$ shown in \cref{fig_P0-z}(\textit{c}), the transverse distributions of cases with $\mathit{Pe}_s=1$ and $2$ are nearly uniform, while the distributions of cases with $\mathit{Pe}_s<1$ still have small fluctuations.
As time continues to increase (not shown here), all the curves will overlap each other and become absolutely uniform \citep{jiang_dispersion_2019}.
The transverse distribution of faster swimmers reaches the uniform equilibrium state much more quickly, as a result of the swimming-induced diffusion.
For $\mathit{Pe}_s=2$, during the transport process, it is clearly observed that the initial high concentration distribution in the middle of the channel decreases fast, resulting in a depletion by the strong swimming effect, as shown in \cref{fig_P0-z}(\textit{b}).
The transport process in other cases is dominated by the comparable effects of the swimming-induce diffusion and translational diffusion.

\begin{figure}
	\centering
	{\includegraphics{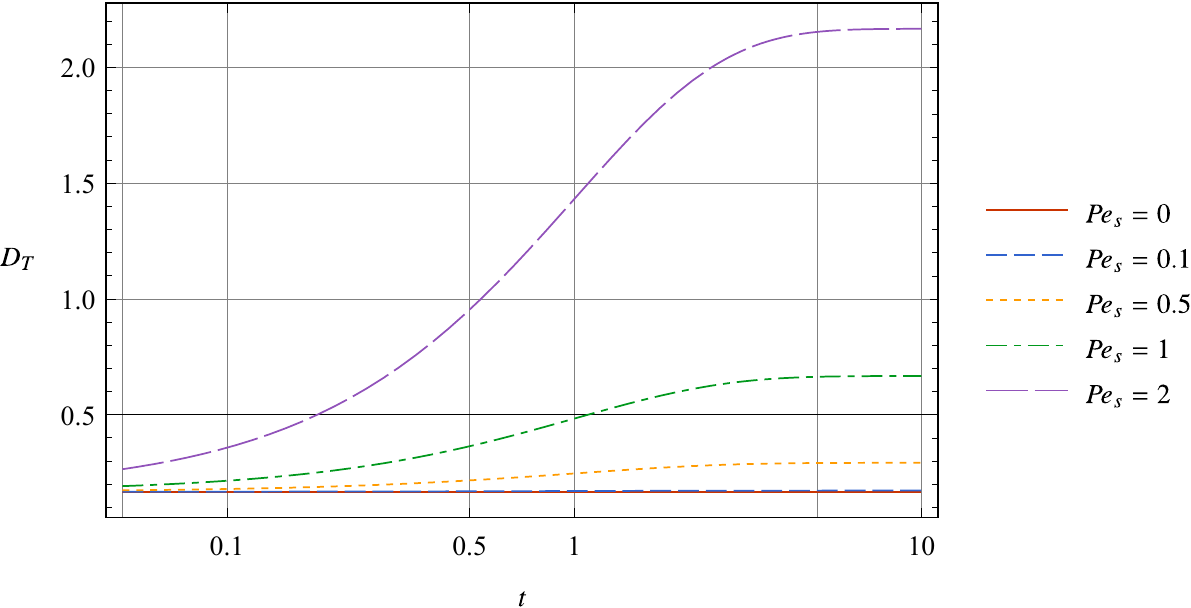}}
	\caption{Temporal evolution of the dispersivity $D_T(t)$ of spherical particles with different swimming ability under the reflective condition without background flow ($\mathit{Pe}_f=0$).
		\label{fig_dispersion}
	}
\end{figure}

\subsubsection{Dispersion characteristics}
\label{sec_results_swimming_dispersion}

Next, we discuss the transient dispersion characteristics related to the moments with order larger than zero.
Note that we do not consider any background flow in this section.
Therefore, the p.d.f.\ of particles is symmetric with respect to the $y$-axis, where the particles are initially released.
Both the drift and the skewness are zero because of this symmetry property.
We only discuss the temporal evolution of the dispersivity.

As shown in \cref{fig_dispersion}, for active particles, the dispersivity increases monotonically with time.
While for passive particles, the dispersivity remains the same as the translational diffusion coefficient, because they only perform pure translation Brownian motions.
In the initial stage of the dispersion process, the dispersivity of active particles, especially those with strong swimming ability (e.g.\ $\mathit{Pe}_s=2$), is quite small.
It then increases rapidly and finally reaches a stable value, i.e.\ the Taylor dispersivity.
Obviously, the larger the $\mathit{Pe}_s$, the larger the dispersivity.
The differences between dispersivities with different $\mathit{Pe}_s$ is gradually enlarged during the transient dispersion process.

Note that without shear flow, the active dispersivity is only comprised of the swimming-induced diffusion (temporal) and the translational diffusion (time-independent).
Actually, in the longitudinal direction, the evolution of the active dispersivity is similar to that of the effective diffusion tensor (the time derivative of the mean squared displacement) in unbounded space \citep{tenhagen_brownian_2011}.
There exists an anomalous dispersion stage before the Taylor dispersion regime. 
Note that when $\mathit{Pe}_s$ is large, the swimming-induced diffusion is the main factor of the dispersivity.
%There is no doubt that the dispersivity depends on the initial condition.
In the initial stage  ($t<0.5$) after the point-source release, the swimming of particles with rotational Brownian motions makes the local distribution uniform in the cross-section, as shown in \cref{fig_2D-P0-phi-theta}(\textit{j},\textit{m},\textit{k},\textit{n}).
Namely, particles can swim randomly at different transverse positions.
The swimming-induced dispersivity in the longitudinal direction is continuously enhanced, which leads to a super-diffusion process.
The enhancement of dispersivity is stopped until the longitudinal length scale of the swimmer cloud is much larger than both the transverse length scale of the cross-section and the length scale  of the swimming range.
The local distribution in the cross-section and the orientation space is nearly uniform at each longitudinal positions, thus the dispersivity finally reaches its maximum value.

\subsection{Influence of shear flow}
\label{sec_results_shear}

We have discussed the swimming effect on the transient dispersion process.
Now we focus on the influence of the shear flow and the combined effect of the shear-induced dispersivity and the swimming-induced diffusion.
To compare with the case without background flow in \cref{sec_results_swimming}, we analyse five cases with different flow P{\'e}clet numbers $\mathit{Pe}_f$ but a fixed swimming P{\'e}clet number $\mathit{Pe}_s=1$.
In the same way, results of spherical particles at three small sample times are plotted to demonstrate the transient process, and only the reflective boundary condition \cref{eq_Pn_reflective_BC} is considered.

\subsubsection{Local distribution: zeroth-order moment}
\label{sec_results_shear_P0}

% Initial stage: a-m

\begin{figure}
	\centering
	{\includegraphics{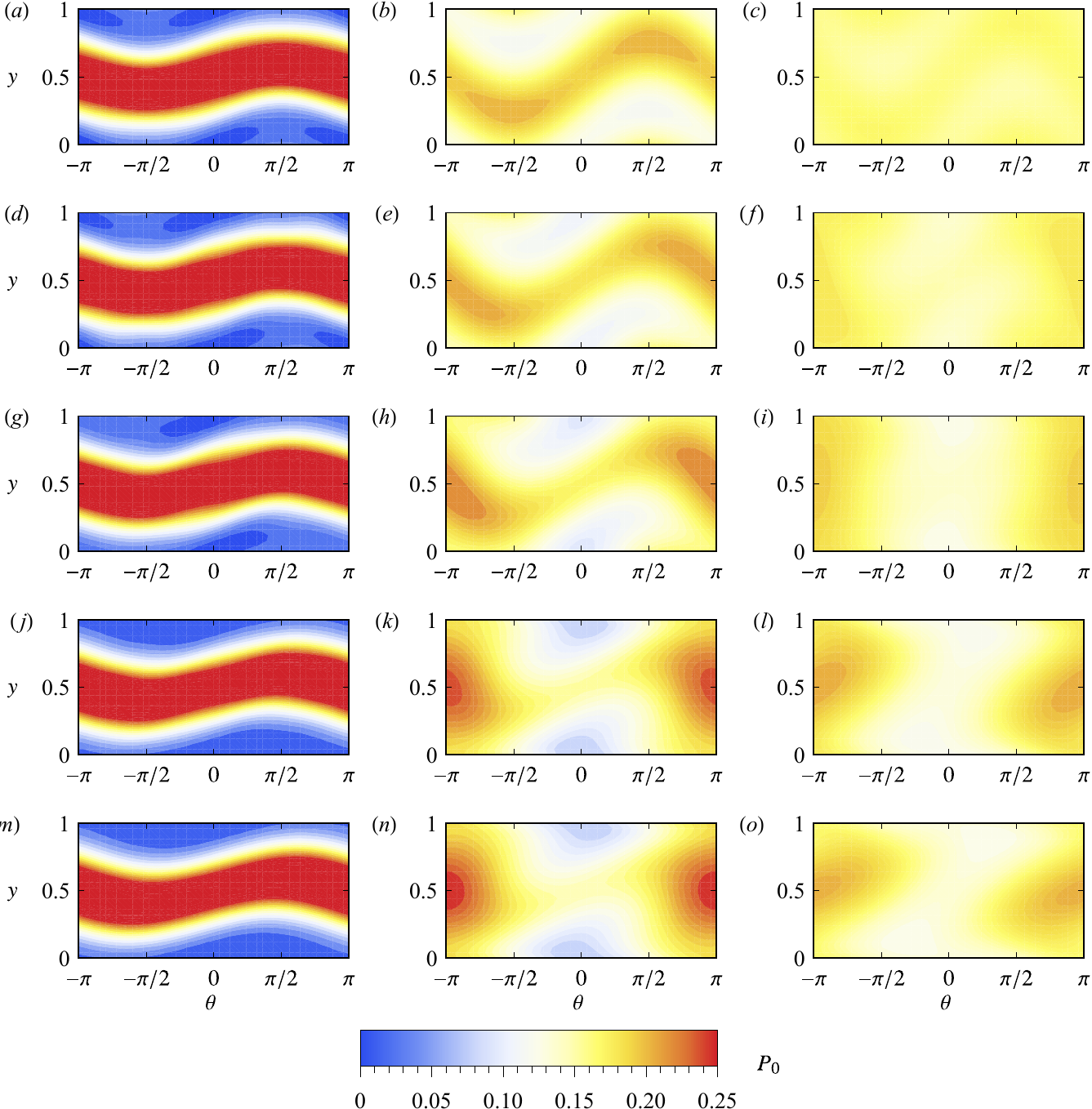}}
	\caption{Density plot of transient local distributions $P_0(y,\theta,t)$ of spherical particles in flows with different flow rates under the reflective condition.
		The flow P{\'e}clet number: 
		(\textit{a}--\textit{c}) $\mathit{Pe}_f =0.1$; (\textit{d}--\textit{f}) $\mathit{Pe}_f =1$; (\textit{g}--\textit{i}) $\mathit{Pe}_f =2$; (\textit{j}--\textit{l}) $\mathit{Pe}_f =4$; (\textit{m}--\textit{o}) $\mathit{Pe}_f =5$.
		Sample times: (\textit{a},\textit{d},\textit{g},\textit{j},\textit{m}) $t=0.1$; (\textit{b},\textit{e},\textit{h},\textit{k},\textit{n}) $t=0.3$;
		(\textit{c},\textit{f},\textit{i},\textit{l},\textit{o}) $t=0.5$.
		In all cases, $\mathit{Pe}_s=1$.
		\label{fig_2D-P0-phi-theta_Flow}
	}
\end{figure}

In the initial stage soon after the point-source release, as shown in \cref{fig_2D-P0-phi-theta_Flow} (\textit{a},\textit{d},\textit{g},\textit{j},\textit{m}) with $t=0.1$, the local distributions of swimmers in the plane Poiseuille flow with different $\mathit{Pe}_f$ are similar. 
The parallel flow carries the swimmers downstream quickly, while the vertical positions of the particles remain unchanged.
Therefore,the swimming diffusion effect is dominant in making the local distribution uniform.
Note that the shear flow can rotate the swimming direction of the particle, which is similar to the rotational Brownian motion, and thus it can also weaken the swimming diffusion effect.
However, in the middle of the channel, the vorticity of the flow is zero.
Therefore, the vorticity-induced rotation is very weak until particles spread over the cross-section of the channel.

% transient stage: b-n
As time increases, unlike the no-flow case discussed in \cref{sec_results_swimming}, the swimmers in a plane Poiseuille flow gradually accumulate at the point  $(y=\frac{1}{2}, \theta=\upi)$ in the local space, as shown in \cref{fig_2D-P0-phi-theta_Flow}(\textit{b},\textit{e},\textit{h},\textit{k},\textit{n}) with $t=0.3$.
Namely, particles mainly swim upstream and near the middle of the channel.
This phenomenon can be explained using the dynamical systems theory.
As discussed in previous studies \citep{zottl_nonlinear_2012,zottl_periodic_2013,jiang_dispersion_2019}, the transverse swimming velocity and the angular velocity can be viewed as a local velocity field in the local space.
For the spherical particles in the plane Poiseuille flow, $(y=\frac{1}{2}, \theta=\upi)$ is a centre point, where particles perform the swing motion around the centreline of the channel \citep{zottl_nonlinear_2012} and closed orbits in the local space are formed.
When the shear is strong, as shown in \cref{fig_2D-P0-phi-theta_Flow}(\textit{k},\textit{n}) with $\mathit{Pe}_f =4, 5$, this temporary accumulation is so intense that the local distribution forms a clear circular spot at $(y=\frac{1}{2}, \theta=\upi)$ (also at $(y=\frac{1}{2}, \theta=-\upi)$ due to the periodicity).

% large times: c-o
At larger times, the local distribution  approaches the uniform distribution, the same as that without background flow discussed in \cref{sec_results_swimming}.
This is also true for any case with a unidirectional flow:
The long-time asymptotic local distribution of spherical swimmers under reflective boundary condition is a uniform distribution \citep{jiang_dispersion_2019}.
The critical point $(y=\frac{1}{2}, \theta=-\upi)$ is only a centre point, which is not stable.
Thus, the accumulation at $(y=\frac{1}{2}, \theta=-\upi)$ dissipates gradually and the local distribution becomes more and more uniform, as shown in \cref{fig_2D-P0-phi-theta_Flow}(\textit{c},\textit{f},\textit{i},\textit{l},\textit{o}) with $t=0.5$.
There is no doubt that with stronger shear, the approach to the homogeneous equilibrium state will be much slower.

\begin{figure}
	\centering
	{\includegraphics{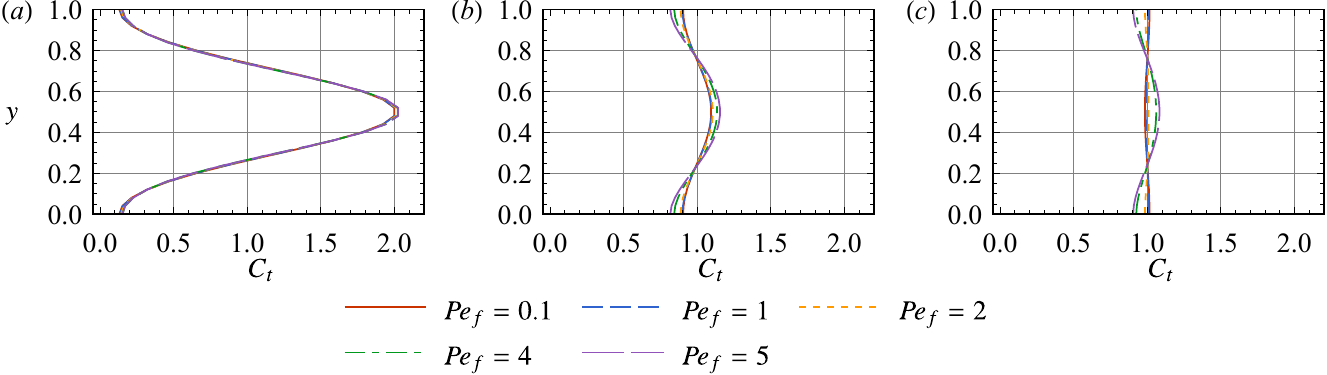}}
	\caption{Transverse distributions $C_t(y,t)$ of spherical particles in flows with different flow rates under the reflective condition.
		Sample times: (\textit{a}) $t=0.1$; (\textit{b}) $t=0.3$;
		(\textit{c}) $t=0.5$.
		In all cases, $\mathit{Pe}_s=1$.
		\label{fig_P0-z_Flow}
	}
\end{figure}

The approach to the uniform distribution in the local space can be demonstrated more clearly with the transverse distribution, defined in \cref{eq_def_transverse_distribution}.
Compared with the case without background flow in \cref{fig_P0-z}, 
there is no concentration depletion  in the middle of the channel, as shown in \cref{fig_P0-z_Flow}(\textit{b}).
Instead, the concentration at $y=\frac{1}{2}$ is the highest, due to the vorticity-induced centre-point accumulation.
When $t=0.5$ as shown in \cref{fig_P0-z_Flow}(\textit{c}), the transverse distribution of swimmers in a low flow rate flow (small $\mathit{Pe}_f$) is nearly uniform.
However, when $\mathit{Pe}_f$ is large enough, e.g.\ $\mathit{Pe}_f=4,5$, there are still observable variations of the transverse distribution from the uniform distribution.
The attenuation of the accumulation is slow and the homogeneous equilibrium state will be reached at larger times (not shown here).
% maybe give a example of the larger time value.

\subsubsection{Dispersion characteristics}
\label{sec_results_shear_dispersion}

Next, we analyse the transient dispersion characteristics.
First, we discuss the drift, i.e.\ the time derivative of the first-order mean concentration moment.
Note that we have transformed the reference to that moving with the mean flow, as in \cref{eq dimensionless variable}.
Thus, the drift discussed here is the average of the longitudinal component of the velocity of particles above the mean flow.

\begin{figure}
	\centering
	{\includegraphics{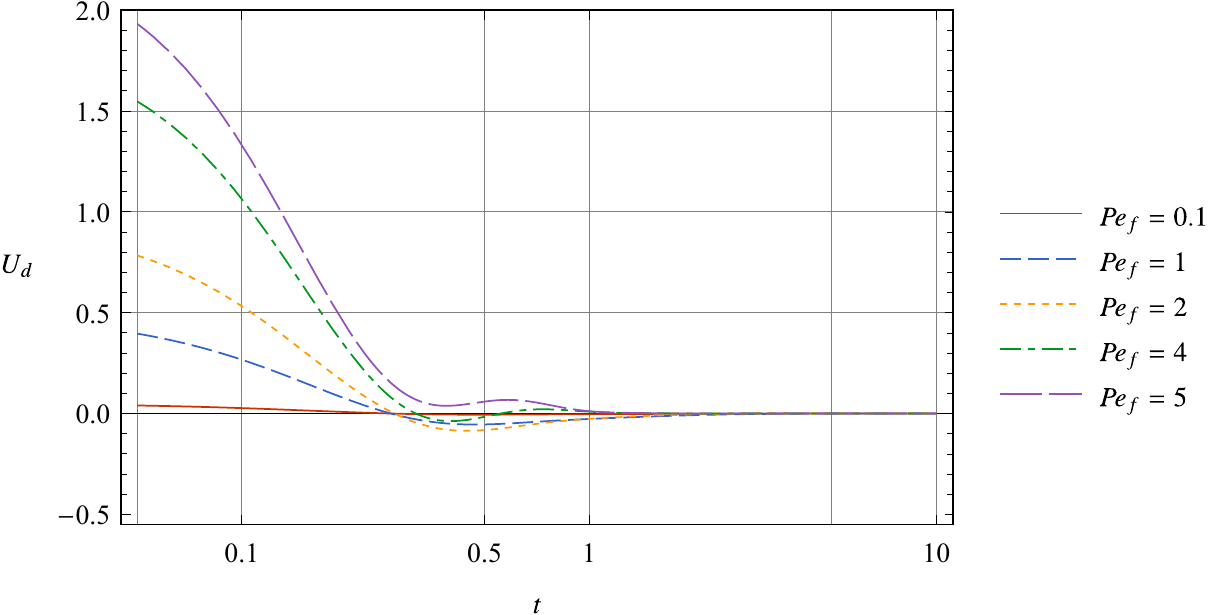}}
	\caption{Temporal evolution of the drift $U_d(t)$ of spherical particles in flows with different flow rates under the reflective condition.
		In all cases, $\mathit{Pe}_s=1$.
		\label{fig_advection_Flow}
	}
\end{figure}

Unlike the case without background flow, the drift of swimmers in a plane Poiseuille flow is not zero in the transient stage, as shown in \cref{fig_advection_Flow}.
In fact, the drift is not small and is positive when the flow rate (represented by $\mathit{Pe}_f$) is large, especially in the initial stage soon after the point-source release in the middle of the channel, where the flow velocity is the largest in the cross-section and thus is larger than the mean flow rate.
Then the drift decreases very fast as time increases.
As shown in \cref{fig_advection_Flow}, all the curves fall to around zero before $t=0.5$.
With a larger $\mathit{Pe}_f$, the initial drift is larger, and thus the decrease rate of the drift is faster.

There are two main factors for the sharp drop of the drift: the advection and the swimming.
First, the spread of particles from the highest-flow-velocity region (in the middle of the channel) to the low-flow-velocity region can reduce the advection velocity of the particles.
Second, due to the swing motion of swimmers in the middle of the channel (centre-point accumulation as shown in \cref{fig_2D-P0-phi-theta_Flow}(\textit{k},\textit{n})), particles mainly swim in the opposite direction of the flow (upstream $\theta=\pm \upi$).
Thus, the corresponding contribution to the drift is negative.
When the swimming effect is dominant (when $\mathit{Pe}_f$ is small), the overall drift can even reduce to below zero, as shown by the curves with $\mathit{Pe}_f=0.1, 1$ in \cref{fig_advection_Flow}.
While the drift curves with larger $\mathit{Pe}_f$, e.g.\ $\mathit{Pe}_f=5$, remain positive during the whole transient stage.

After the sharp drop, the overall drift slightly increases with time, for all the curves in \cref{fig_advection_Flow}.
Because the local distribution become more and more uniform as time increases, as shown in \cref{fig_2D-P0-phi-theta_Flow},
the upstream swimming effect is weakened and the reduction of the drift is partly recovered.
At larger times ($t>1$), all the drift curves approach zero.
Because the long-time asymptotic local distribution is uniform, the corresponding overall drift by \cref{eq_moment_governing_Mn} is
\begin{equation*}
	\lim_{t \rightarrow \infty} U_d = \lim_{t \rightarrow \infty}
	\overline{\left( \mathit{Pe}_f u + \mathit{Pe}_s \cos \theta \right) P_0} =
	\overline{\left( \mathit{Pe}_f u + \mathit{Pe}_s \cos \theta \right)} = 0,
\end{equation*}
as discussed by our previous study \citep{jiang_dispersion_2019}.
The mass centre of the swimmer cloud finally moves with the mean flow rate.
However, there are great differences among the approach-to-zero process of the drift with different $\mathit{Pe}_f$.
For small $\mathit{Pe}_f=1, 2$, $U_d$ increases to zero directly from the lowest negative value caused by the sharp drop.
For larger $\mathit{Pe}_f=4, 5$, $U_d$ increases slowly for a while to some positive values, and finally decreases to zero.
The curve with $\mathit{Pe}_f=4$ shows a fluctuation across the zero value, while the drift with $\mathit{Pe}_f=5$ remains positive,
as a result of the complex combined reduction effect of the advection and the swimming.

Next, we discuss the temporal evolution of dispersivity.
In \cref{sec_results_swimming_dispersion}, the dispersivity is only compromised of the swimming-induced diffusion and the translational diffusion.
Adding the effect of the shear flow makes the evolution of the comprehensive dispersivity much more complicated.
The overall dispersivity is not a simple superposition of the shear-enhanced dispersivity and the swimming-induced diffusion.
In fact, the shear effect and the swimming effect can inhibit each other!
To analyse the overall dispersivity, one should bear in mind the question that which effect is dominant.

\begin{figure}
	\centering
	{\includegraphics{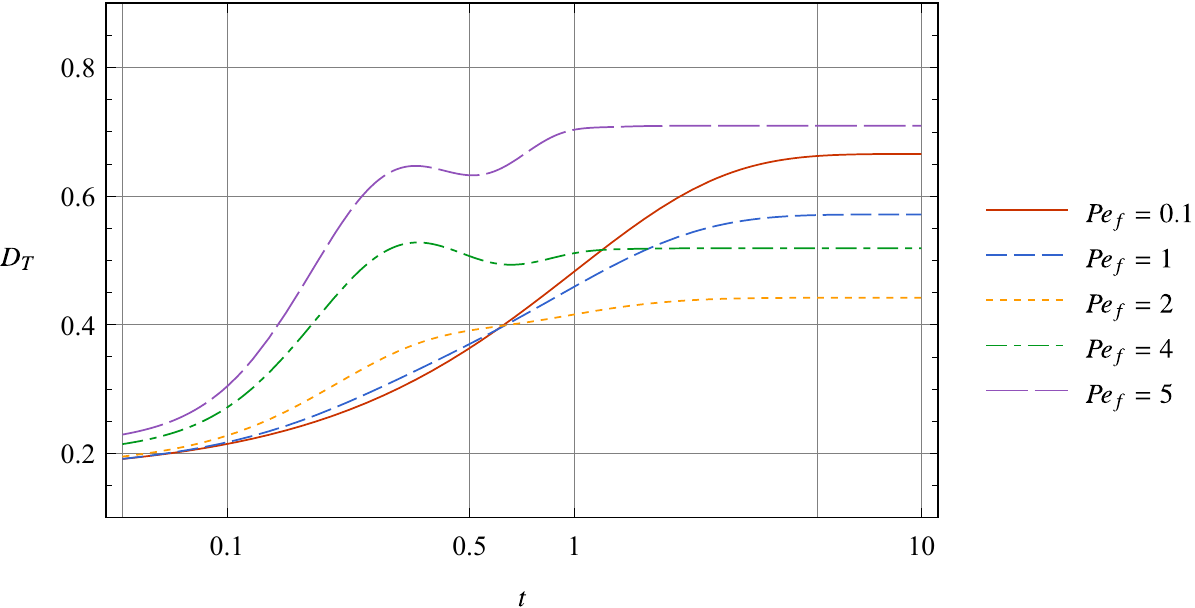}}
	\caption{Temporal evolution of the dispersivity $D_T(t)$ of spherical particles in flows with different flow rates under the reflective condition.
		In all cases, $\mathit{Pe}_s=1$.
		\label{fig_dispersion_Flow}
	}
\end{figure}

When $\mathit{Pe}_f$ is small, the swimming-induced diffusion is dominant in the dispersion process.
As shown in \cref{fig_dispersion_Flow}, the curves with $\mathit{Pe}_f=0.1, 1, 2$ are similar to that without background flow in \cref{fig_dispersion}: 
the overall dispersivity increases monotonically with time.
In the initial stage ($t<0.5$), the dispersivities with larger $\mathit{Pe}_f=1,2$ are larger and increase faster than that with $\mathit{Pe}_f=0.1$.
Because the transverse distribution becomes more uniform due to the swimming, as shown in \cref{fig_P0-z_Flow}, the shear-enhanced dispersivity becomes stronger as the particles spread from the low shear-rate region (the middle of the channel) to the high shear-rate regions.
The distribution of the swimming direction is still highly non-uniform, and thus the increase of the swimming-induced dispersivity is slow.
However, at large times ($t>1$), the increases of the dispersivities with larger $\mathit{Pe}_f=1,2$ become smaller.
More importantly, the long-time asymptotic values, i.e.\ the Taylor dispersivities, are much smaller than that with $\mathit{Pe}_f=0.1$ \citep{jiang_dispersion_2019}.
Note that at large times, the swimming-induced diffusion gradually exerts its influence and regains the dominance in the dispersion process,  as the whole local distribution become much more uniform.
The shear-enhanced rotation of the swimming direction can weaken the swimming-induced diffusion, as discussed in \cref{sec_results_shear_P0}.
Therefore, with larger $\mathit{Pe}_f=1,2$, the Taylor dispersivities dominated by the swimming-induced diffusion are smaller.

For large $\mathit{Pe}_f$, as shown by the curves with $\mathit{Pe}_f=4, 5$ in \cref{fig_dispersion_Flow},
the evolution of the dispersivity is more complex and does not monotonically increase with time.
Note that the shear-enhanced dispersivity by advection is dominant.
Thus, there is a very rapid rise of the dispersivity in the initial transient stage ($t<0.5$), which is similar to the case with low $\mathit{Pe}_f$.
It is followed by an obvious but small reduction of the dispersivity, as a result of the inhibition by the swimming-induced diffusion.
Note that in the case of passive particles, in the shear-dominant dispersion regime, increasing the translational diffusion will decrease the Taylor dispersivity (see equation (41) in the work of \cite{aris_dispersion_1956}).
Similarly, the swimming-induced diffusion can also suppress the shear dispersion \citep{bearon_spatial_2011,jiang_dispersion_2020}.
Finally, the dispersivity increases with time again  and reaches the equilibrium state.
For $\mathit{Pe}_f=4$, the  long-time asymptotic value is smaller than the maximum value and that of the case without flow, as a result of the mutual inhabitation of the shear dispersion between the swimming-induced diffusion.
For $\mathit{Pe}_f=5$, the shear dispersion achieves absolute dominance: the finial value exceeds that without flow which is compromised only by the swimming-induced diffusion and translational diffusion.

\begin{figure}
	\centering
	{\includegraphics{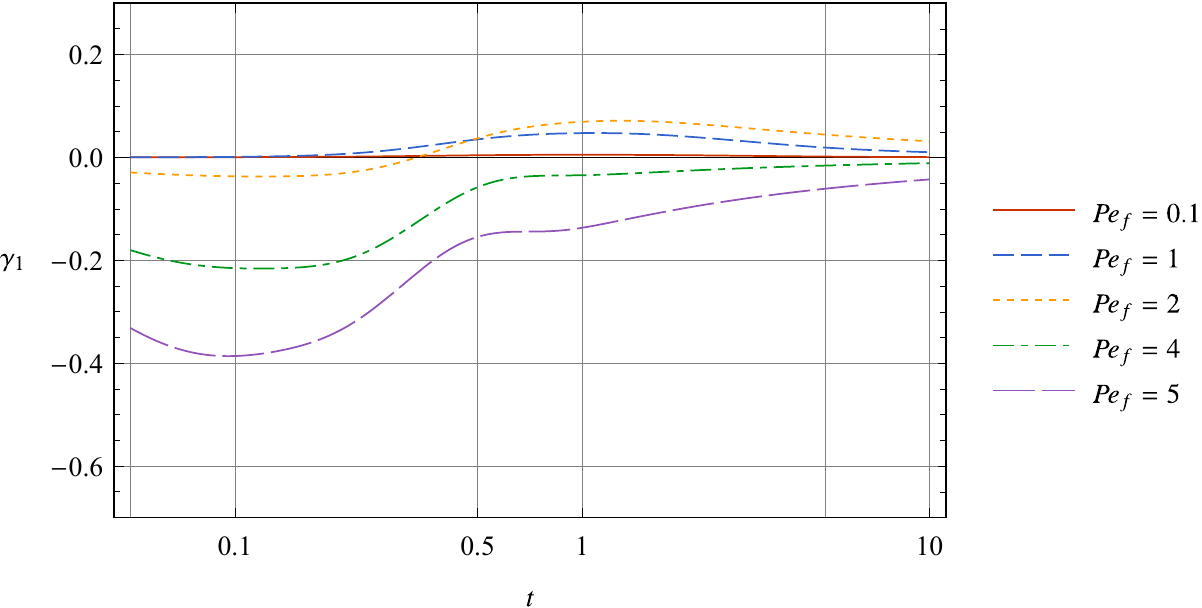}}
	\caption{Temporal evolution of the skewness $\gamma_1(t)$ of spherical particles in flows with different flow rates under the reflective condition.
		In all cases, $\mathit{Pe}_s=1$.
		\label{fig_skew_Flow}
	}
\end{figure}

Finally, we discuss the skewness caused by the shear flow.
As shown in \cref{fig_skew_Flow}, the temporal evolution of skewness is much more complicated than those of the drift and dispersivity.
Similar to the case of passive particles \citep{aris_dispersion_1956,aminian_how_2016},  the skewness is negative  in the initial transient stage, as a result of the dominant advection effect by the plane Poiseuille flow.
When $\mathit{Pe}_f$ is small (e.g.\ $\mathit{Pe}_f=0.1, 1$), the swimming-induced diffusion effect is stronger than the advection effect.
The skewness is small and negative in the initial stage, and then becomes positive as time increases.
Note that the skewness under the pure swimming-induced diffusion is zero, as discussed in \cref{sec_results_swimming_dispersion}.
Thus, the positive skewness is due to the combined effect of the swimming-induced diffusion and the advection, more specifically, by the vorticity-induced rotation of the swimming directions of particles.
For a plane Poiseuille flow, the vorticity-induced rotation is strong near the wall where the shear rate is large.
Therefore, the cloud of particles in the middle of the channel travelling downstream disperses faster than that near the walls travelling upstream (relative to the mean flow rate), due to the swimming induced diffusion. 
The downstream part of the mean distribution is more uniform in the longitudinal direction, which results in the positiveness of the skewness.

For the cases with large $\mathit{Pe}_f$ (e.g.\ $\geqslant 2$), the advection effect is dominant.
The negativeness of the skewness is obviously observed and the temporal variation of skewness is large at small times.
The skewness first decreases as time increases ($0.1<t<0.5$), due to the advection effect.
Then it greatly increases, because of the comprehensive combined effect of the swimming-induced diffusion and the advection.

Finally, at large times, the skewness gradually approaches zero, for all the cases in \cref{fig_skew_Flow}.
This means that the asymmetry of the mean concentration distribution disappears and indicates that the distribution becomes Gaussian.
The approach to zero (or to the Taylor dispersion regime) is very slow.
Even when the dispersivity reaches its equilibrium value (about $t>5$), there is still small varying skewness of the mean concentration distribution.

\subsection{Influence of boundaries: wall accumulation}

The above-discussed cases are under the reflective boundary condition \cref{eq_Pn_reflective_BC}.
Now we turn to the Robin condition \cref{eq_Pn_Robin_BC} to consider the influence of wall accumulation on the transient dispersion process of spherical particles.
To demonstrate the combined effect of wall accumulation with the shear flow and the swimming-induced diffusion, we choose six cases, with the flow P{\'e}clet numbers $\mathit{Pe}_f\in \{0.1, 2, 5\}$ and the swimming P{\'e}clet numbers $\mathit{Pe}_s\in \{0.1, 1\}$.
The same three sample times are chosen to compare with the results without accumulation.

\subsubsection{Local distribution: zeroth-order moment}
\label{sec_results_P0_Robin}

\begin{figure}
	\centering
	{\includegraphics{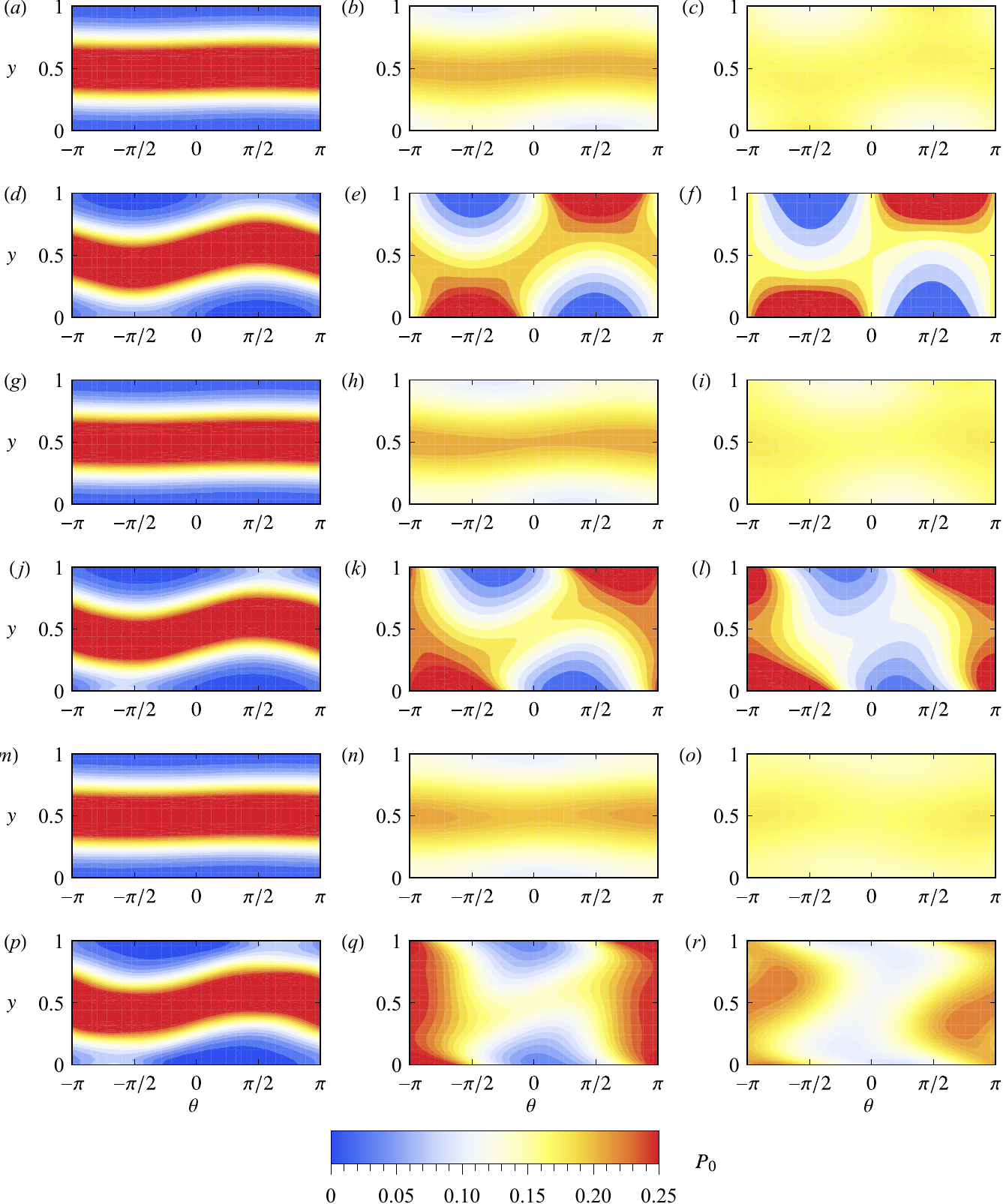}}
	\caption{Density plot of transient local distributions $P_0(y,\theta,t)$ of spherical particles under the Robin condition.
		The P{\'e}clet numbers:
		(\textit{a}--\textit{c}) $\mathit{Pe}_s=0.1, \mathit{Pe}_f=0.1$; (\textit{d}--\textit{f}) $\mathit{Pe}_s=1, \mathit{Pe}_f=0.1$; (\textit{g}--\textit{i}) $\mathit{Pe}_s=0.1, \mathit{Pe}_f=2$; (\textit{j}--\textit{l}) $\mathit{Pe}_s=1, \mathit{Pe}_f=2$;
		(\textit{m}--\textit{o}) $\mathit{Pe}_s=0.1, \mathit{Pe}_f=5$; (\textit{p}--\textit{r}) $\mathit{Pe}_s=1, \mathit{Pe}_f=5$.
		Sample times: (\textit{a},\textit{d},\textit{g},\textit{j},\textit{m},\textit{p}) $t=0.1$; (\textit{b},\textit{e},\textit{h},\textit{k},\textit{n},\textit{q}) $t=0.3$;
		(\textit{c},\textit{f},\textit{i},\textit{l},\textit{o},\textit{r}) $t=0.5$.
		\label{fig_2D-P0-phi-theta_Robin}
	}
\end{figure}

\begin{figure}
	\centering
	{\includegraphics{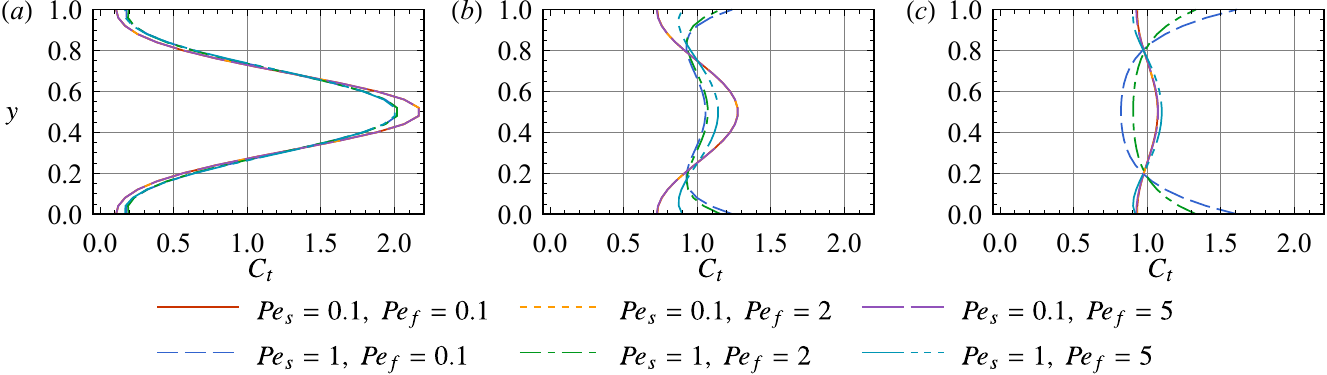}}
	\caption{Transverse distributions $C_t(y,t)$ of spherical particles under the Robin condition.
		Sample times: (\textit{a}) $t=0.1$; (\textit{b}) $t=0.3$;
		(\textit{c}) $t=0.5$.
		\label{fig_P0-z_Robin}
	}
\end{figure}

% the temporal evolution
As shown in \cref{fig_2D-P0-phi-theta_Robin}, there are fundamental differences between the local distribution under the Robin condition and that in \cref{fig_2D-P0-phi-theta_Flow} under the reflective condition.
At the very initial stage after the point-source release, the local distributions are similar under these two types of condition, mainly depending on the swimming ability  ($\mathit{Pe}_s$).
As swimmers reach the wall, they gradually form an obvious and sustained accumulation among the incoming angle range (e.g.\ $-\upi<\theta<0$ at the wall $y=0$).
Under the Robin condition \cref{eq_Pn_Robin_BC}, there is no penetration of particles through the walls in the phase space, for each swimming angle.
Therefore, particles can only change their swimming direction by rotational diffusion.
The incoming swimming probability flux is balanced by the translational flux with a negative wall-normal concentration gradient, as clearly shown in \cref{fig_2D-P0-phi-theta_Robin}(\textit{e},\textit{k}) at $t=0.3$ with $\mathit{Pe}_s=1$.
Meanwhile, for the outgoing swimming angle ($0<\theta<\upi$), the value of the distribution is very small and a positive wall-normal concentration gradient is established at the walls.
Taken together, particles mainly swim towards the walls and thus accumulate at the walls.
At larger times, the wall-accumulated distribution by the incoming flux of particles 
remains and does not approach a uniform distribution as that under the reflective condition \citep{ezhilan_transport_2015,jiang_dispersion_2019}.
Namely, the equilibrium state of the local transport under the Robin condition is not homogeneous.
The wall accumulation process can be demonstrated more clearly using the transverse distribution, as defined in \cref{eq_def_transverse_distribution} and shown in \cref{fig_P0-z_Robin}.

% Parameters
Comparing the local distribution with different $\mathit{Pe}_s$ and $\mathit{Pe}_f$ in \cref{fig_2D-P0-phi-theta_Robin,fig_P0-z_Robin},
the wall accumulation is enhanced by stronger swimming ability but is suppressed by the shear flow.
When $\mathit{Pe}_s=1$, the accumulation strength and the incoming-angle-preferred orientation distribution are completely distinct from those with $\mathit{Pe}_s=0.1$.
The stronger the incoming swimming probability flux, the larger the wall-normal concentration gradient, and thus the stronger the wall accumulation.
As for the influence of the shear flow, when $\mathit{Pe}_f$ is large enough and the vorticity-induced rotation is strong, as shown in \cref{fig_2D-P0-phi-theta_Robin}(\textit{q},\textit{r}) and \cref{fig_P0-z_Robin}(\textit{c}), the wall accumulation is greatly weakened and even disappears.
Additionally, the incoming-angle-preferred distribution of $\theta$ remains but 
is nearly confined to only the half of the range, e.g.\ $-\upi<\theta<-\upi/2$ at the wall $y=0$.
As discussed in \cref{sec_results_shear_P0} and previous studies \citep{zottl_nonlinear_2012,jiang_dispersion_2019},
the vorticity-induced swing motions of particles around the centreline of the channel lead to the centre-point accumulation in the local space, which compensates the centreline depletion by the Robin condition.
Furthermore, particles mainly swim upstream (parallel to the streamline).
Thus, the incoming flux is weakened, reducing the strength of the wall accumulation.

\subsubsection{Dispersion characteristics}
\label{sec_res_Robin_dispersion}

\begin{figure}
	\centering
	{\includegraphics{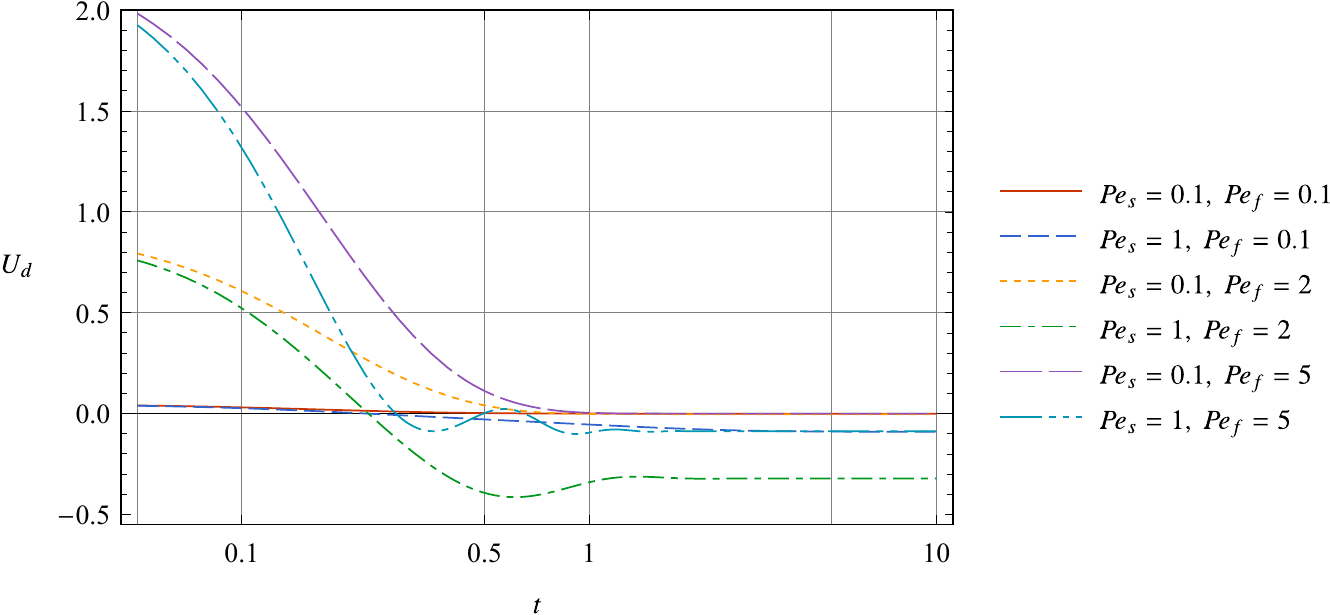}}
	\caption{Temporal evolution of the drift $U_d(t)$ of spherical particles under the Robin condition.
		\label{fig_advection_Robin}
	}
\end{figure}

Next, we discuss the dispersion characteristics under the Robin condition.
First, for the drift, as shown in \cref{fig_advection_Robin}, there is a sharp drop in the very initial stage, similar to the result shown in \cref{fig_advection_Flow} under the reflective condition.
The advection and the swimming are the two key factors for the drift drop, as discussed in \cref{sec_results_shear_dispersion}.
For the current case, the accumulation is the third main contributor.
Near the walls,  the flow speed relative to the mean flow rate is negative.
Thus, the growing accumulation of particles at the walls drives them to move upstream, which can greatly decrease the drift, the local-distribution-weighted average of the longitudinal component of velocity, as shown in \cref{eq_moment_governing_drift}.
At large times, the wall-accumulation-reduced drift even becomes negative, which is fundamentally different from the case under the reflective boundary condition.
The reason is that the equilibrium state of the local distribution under the Robin condition is not homogenous, as discussed in the previous study \citep{jiang_dispersion_2019}.

% parameter; Approach process.
As shown in \cref{fig_advection_Robin}, the initial value of the drift is highly related to $\mathit{Pe}_f$, which indicates that the advection effect is dominant for the drift in the very initial stage.
The decrease of the drift can be non-monotonic, especially when both $\mathit{Pe}_f$ and $\mathit{Pe}_s$ are large.
For example, the drift curves with $\mathit{Pe}_s=1, \mathit{Pe}_f=1$ and $\mathit{Pe}_s=1, \mathit{Pe}_f=5$ rise slightly after the rapid drop, 
which is similar to the case under the reflective boundary condition, as discussed in \cref{sec_results_shear_dispersion}.
The long-time asymptotic mainly depends on $\mathit{Pe}_s$ because the swimming ability mainly determines the strength of the wall accumulation, as discussed in \cref{sec_results_shear_P0}.
With small $\mathit{Pe}_s=0.1$, the wall accumulation is weak, thus the equilibrium drift is nearly zero for all the cases with different $\mathit{Pe}_f$.
With $\mathit{Pe}_s=1$, the equilibrium drift is negative and far from zero due to the strong wall accumulation.

\begin{figure}
	\centering
	{\includegraphics{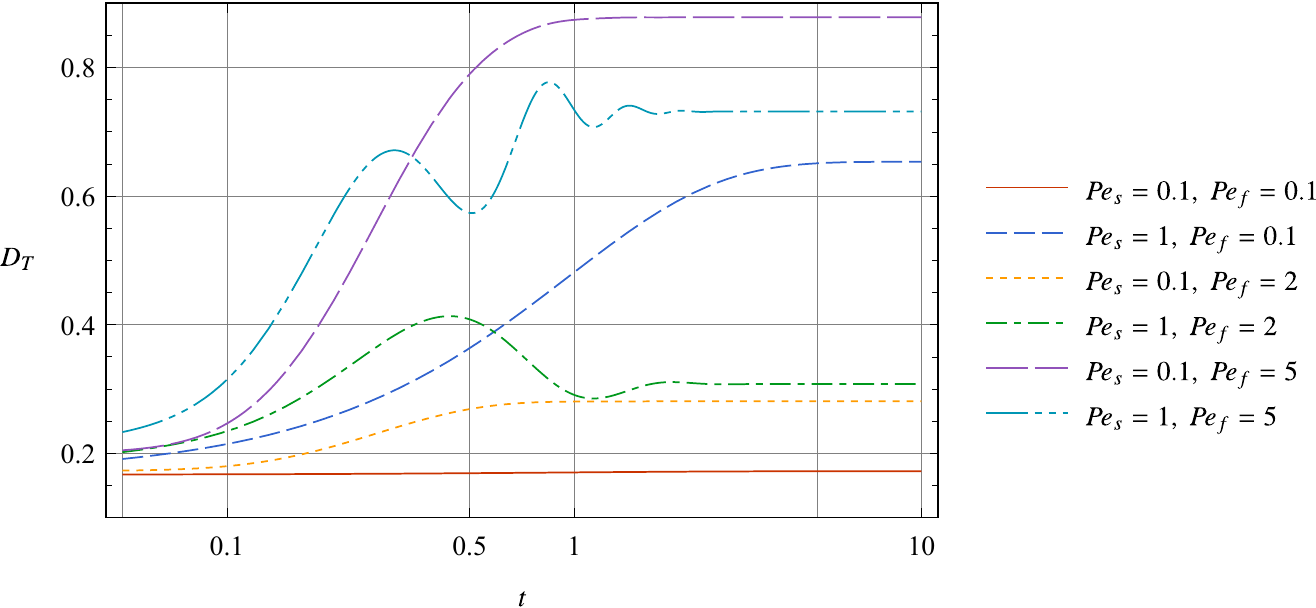}}
	\caption{Temporal evolution of the dispersivity $D_T(t)$ of spherical particles under the Robin condition.
		\label{fig_dispersion_Robin}
	}
\end{figure}

Now we turn to the temporal evolution of the dispersivity.
As shown in \cref{fig_dispersion_Robin}, there is an overall upward trend of the dispersivity, from a small initial value to the larger Taylor dispersivity, which is similar to the case under the reflective boundary condition discussed in \cref{sec_results_shear_dispersion}.
In the very initial stage, the wall accumulation is not fully formed because most particles are still far away from the walls after the point-source release, as discussed in \cref{sec_results_P0_Robin}.
Therefore, the increase of the dispersivity is the combined result of the shear dispersion, swimming-induced diffusion and translational diffusion.

As particles spreads toward the walls, the wall accumulation exerts its influence, especially for the cases with large values of both $\mathit{Pe}_s$ and $\mathit{Pe}_f$.
Comparing the curve with $\mathit{Pe}_s=1$ and $\mathit{Pe}_f=2$ in \cref{fig_dispersion_Robin} with that in \cref{fig_dispersion_Flow} under the reflective condition, 
the accumulation makes the dispersivity decrease earlier (around $t=0.5$) and more considerably.
The dispersivity also experiences a slight rise after the drop, but finally approaches a smaller equilibrium value.
As discussed in our previous study \citep{jiang_dispersion_2019}, the wall accumulation can suppress the dispersion process in the plane Poiseuille flow, for both the swimming and advection effects.
In the accumulation layer, particles mainly swim towards the wall, and thus the swimming-induced diffusion is weakened.
On the other hand, particles accumulate near the low-flow-speed regions, and thus the advection effect by the relative velocity difference in the cloud of particles is also reduced.
For the curve with $\mathit{Pe}_s=1$ and $\mathit{Pe}_f=5$, the combined effect of the shear dispersion and the wall accumulation is much more complex.
The curve shows strong fluctuations in the transient stage (e.g.\ $0.3<t<1$).
Note that the whole dispersion process is dominant by the advection effect. 
The strength of the wall accumulation is weakened, as discussed in \cref{sec_results_P0_Robin}, 
thus the suppression of the dispersivity is very weak in the Taylor dispersion regime at large times.

\begin{figure}
	\centering
	{\includegraphics{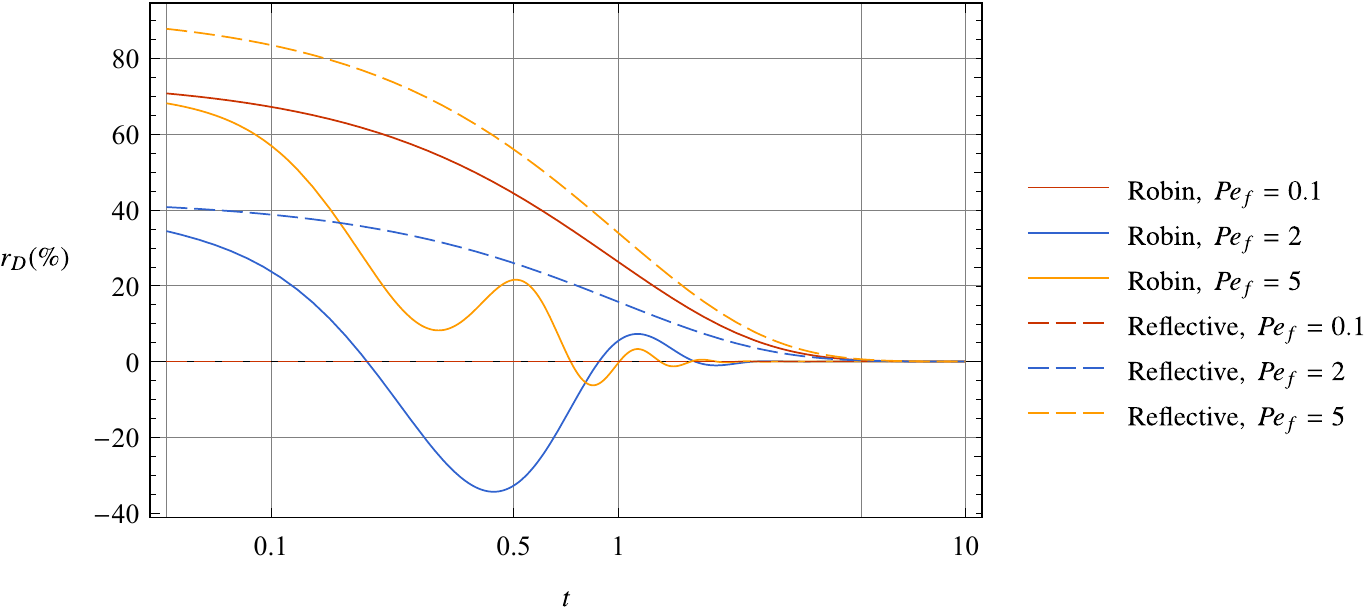}}
	\caption{Temporal evolution of the relative percentage difference $r_D(t)$ of dispersivity for spherical particles under different boundary conditions.
		`Robin' denotes the Robin condition, and `Reflective' denotes the reflective condition.
		In all cases, the swimming P{\'e}clet number $\mathit{Pe}_s=1$.
		\label{fig_dispersion_compare}
	}
\end{figure}

It is of great interest to investigate whether the wall accumulation can slow down or accelerate the approach process to the Taylor dispersion regime, compared with the no-accumulation result under the reflective condition in \cref{sec_results_shear_dispersion}.
To estimate the time scale before entering the Taylor dispersion regime,
we introduce the relative percentage difference of dispersivity: 
\begin{equation} \label{eq_def_relative_difference}
	r_D (t) \triangleq \frac{D_T (t) - D_T^{\infty}}{D_T^{\infty}} \times 100 \%,
\end{equation}
where $D_T^{\infty} \triangleq \lim_{t \rightarrow \infty} D_T(t)$ is the Taylor dispersivity.
A zero $r_D$  indicates that the Taylor regime is reached.

As shown in \cref{fig_dispersion_compare}, comparing the results under the Robin condition and the reflective condition, the wall accumulation slightly influences the time scale for the Taylor regime.
When $\mathit{Pe}_f$ is small (e.g.\ $\mathit{Pe}_f=0.1$), the curves of $r_D$ are nearly the same and the Taylor regime is reached when $t\approx 5$, though the local distributions under these two boundary conditions are fundamentally different, as shown in \cref{fig_2D-P0-phi-theta_Flow}(\textit{c}) and \cref{fig_2D-P0-phi-theta_Robin}(\textit{f}).
Note that when the flow rate is small, the swimming-induced diffusion is dominant in the dispersion process.
The difference between the Robin condition and the reflective condition is whether to change the direction of the vertical motion after a particle hits a wall, as discussed in \cref{sec_simulations}.
However, the direction of the  longitudinal motion remains unchanged under both conditions.
Therefore, the overall longitudinal dispersion process is similar under these two conditions.
When $\mathit{Pe}_f$ is larger (e.g.\ $\mathit{Pe}_f=2$), though the temporal variations of $r_D$ are quite different under these two boundary conditions, they approach zero nearly at the same time ($t<5$).
Only when $\mathit{Pe}_f$ is very large (e.g.\ $\mathit{Pe}_f=5$), the wall accumulation can, to some extent, hinder the dispersion process: there is still small fluctuation of dispersivity under the Robin condition when the dispersivity under the reflective condition is nearly steady.

\begin{figure}
	\centering
	{\includegraphics{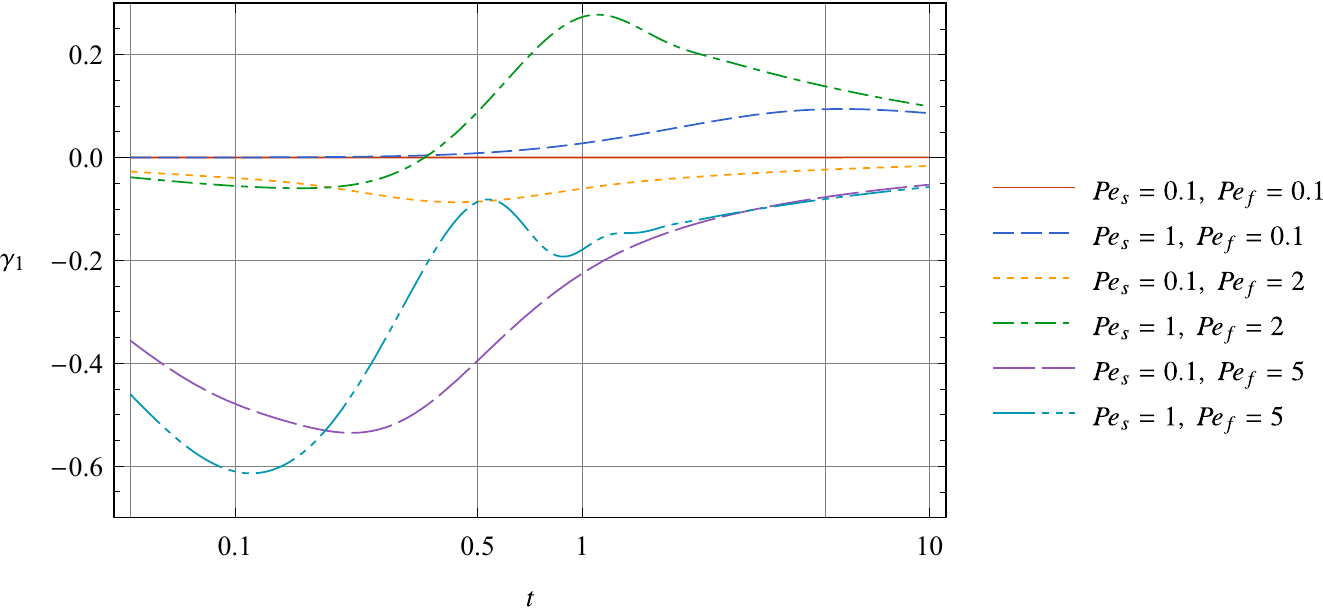}}
	\caption{Temporal evolution of the skewness $\gamma_1(t)$ of spherical particles under the Robin condition.
		\label{fig_skew_Robin}
	}
\end{figure}

Finally, we discuss the temporal evolution of the skewness.
Overall, the wall accumulation can enhance the skewness for both the dispersion regimes dominated by the swimming-induced diffusion and the advection.
First, in the initial stage, the evolution of the skewness is similar to that under the reflective boundary condition.
Comparing \cref{fig_skew_Robin} with \cref{fig_skew_Flow},
the skewness is negative, due to the advection effect.
It first decreases and then increases as time increases.
When $\mathit{Pe}_f$ is small, the swimming-induced diffusion effect is dominant in the dispersion process.
The negative skewness rises and becomes positive at larger times, because of the strong vorticity-induced rotation of the swimming directions of particles near the walls, as discussed in \cref{sec_results_shear_dispersion} for the reflective boundary condition.
Therefore, under the Robin condition, the wall accumulation makes the positive skewness larger.
Much more particles concentrated near the walls and disperse slower than those near the centreline of the channel.
When $\mathit{Pe}_f$ is large and the advection effect is dominant in the dispersion process, the absolute value of the skewness under the Robin condition is larger than that under the reflective condition.
This is because the shear-enhanced dispersivity is larger near the walls where the shear rate is larger for the plane Poiseuille flow.
The wall accumulation can thus strengthen the advection effect.
At large times, the skewness gradually approaches zero for all the cases, which is similar to that under the reflective condition, though the decay process under the Robin condition is slower.

\subsection{Influence of particle shape: shear-induced alignment}
\label{sec_results_shape}

The above discussion considers only the spherical particles.
Now we focus on the general case of ellipsoidal particles.
Unlike spherical particles, ellipsoidal particles (with shape factor $\alpha_0 > 0$) experience not only the rotation induced by the vorticity of the fluid but also the alignment induced by the strain motion of the fluid \citep{ezhilan_transport_2015}, as shown by Jeffery's equation \labelcref{eq angular velocity} for the angular velocity.
For infinitely thin rod-like particles (with $\alpha_0=1$), the shear-induced alignment makes them swim parallel to the streamlines, and thus is also called streamwise alignment and known as the behaviour of rheotaxis \citep{pedley_hydrodynamic_1992}.
To demonstrate the effect of shear-induced alignment and compare with the above-discussed cases, we choose four cases of ellipsoidal particles with $\alpha_0 \in \{0.5, 1\}$ under the Robin condition and reflective boundary condition.
Other parameters are fixed or kept the same with the previous cases:  the swimming P{\'e}clet number $\mathit{Pe}_s = 1$, and the flow P{\'e}clet number $\mathit{Pe}_f =2$.

\subsubsection{Local distribution: zeroth-order moment}
\label{sec_results_P0_shape}

\begin{figure}
	\centering
	{\includegraphics{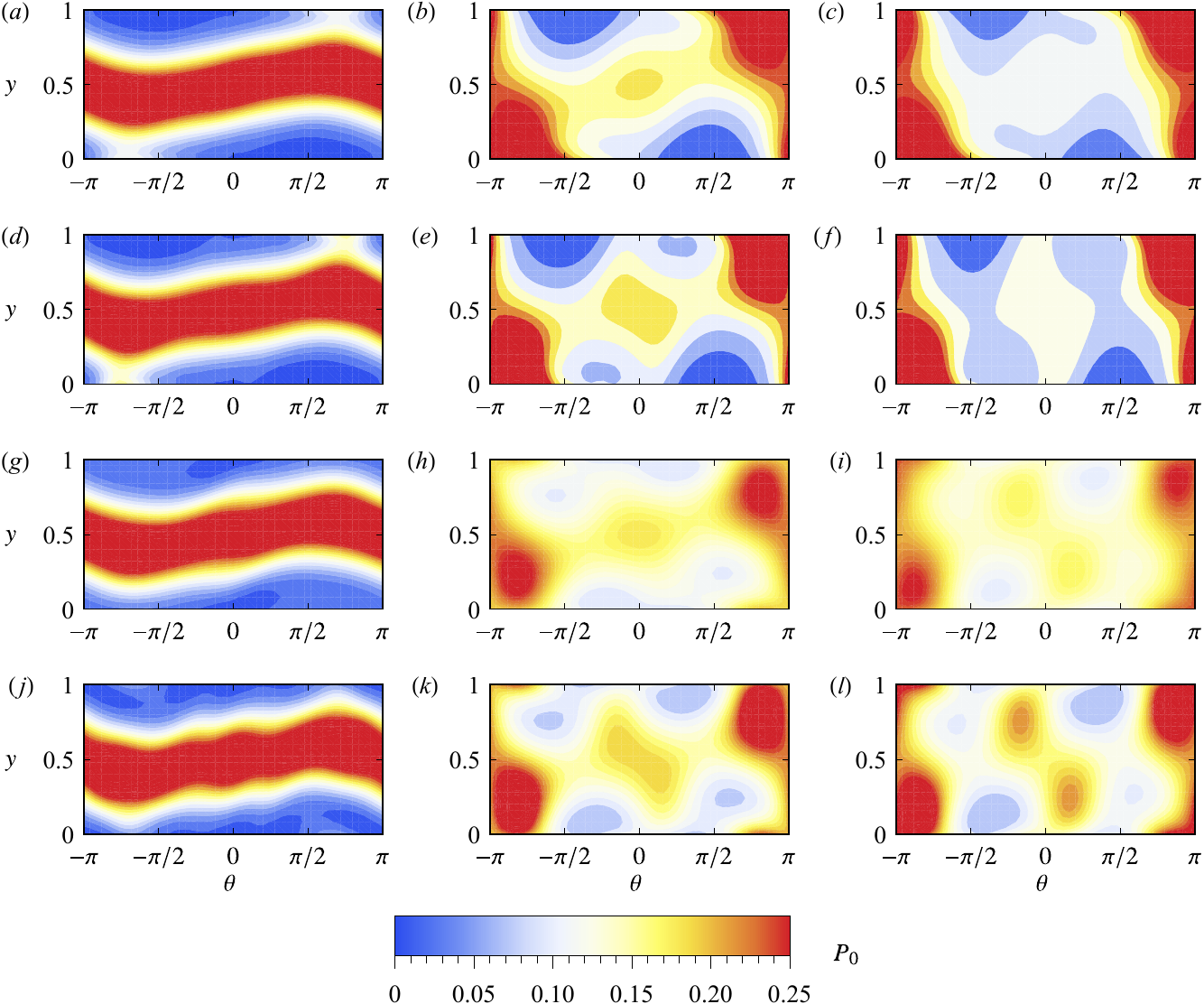}}
	\caption{Density plot of transient local distributions $P_0(y,\theta,t)$ of ellipsoidal particles under the Robin and reflective condition.
		(\textit{a}--\textit{c}) $\alpha_0=0.5$, Robin condition;
		(\textit{d}--\textit{f}) $\alpha_0=1$; Robin condition; 
		(\textit{g}--\textit{i}) $\alpha_0=0.5$, Robin condition; 
		(\textit{j}--\textit{l}) $\alpha_0=1$, reflective condition.
		Sample times: (\textit{a},\textit{d},\textit{g},\textit{j}) $t=0.1$; 
		(\textit{b},\textit{e},\textit{h},\textit{k}) $t=0.3$;
		(\textit{c},\textit{f},\textit{i},\textit{l}) $t=0.5$.
		In all cases, $\mathit{Pe}_s = 1$, $\mathit{Pe}_f=2$.
		\label{fig_2D-P0-phi-theta_shape}
	}
\end{figure}

As shown in \cref{fig_2D-P0-phi-theta_shape}, the shear-induced alignment of ellipsoidal particles significantly affects the distribution of the swimming direction during the transient dispersion process.
First, for the Robin condition, it has been shown in \cref{fig_2D-P0-phi-theta_Robin}(\textit{j},\textit{k},\textit{l}) for spherical particles that the vorticity-induced rotation confines the  incoming-angle-preferred distribution to nearly only the half of the range:
particles mainly swim upstream near the walls ($\theta=\pm \upi$).
The strain-induced alignment further enhances the upstream-preferred angle distribution.
Additionally, some ellipsoidal particles near the walls can swim downstream, which is not observed in the spherical case.

Second, for the reflective condition, the vorticity-induced tendency of upstream swimming of spherical particles in the middle of the channel after the release is weakened for ellipsoidal particles.
Comparing \cref{fig_2D-P0-phi-theta_shape}(\textit{g}--\textit{l}) with \cref{fig_2D-P0-phi-theta_Flow}(\textit{g},\textit{h},\textit{i}),
ellipsoidal particles near the walls mainly swim upstream, the same as those under the Robin condition.
While in the middle of the channel, some particles swim downstream, due to the shear-induced alignment effect, which is  different from the spherical particles.

\begin{figure}
	\centering
	{\includegraphics{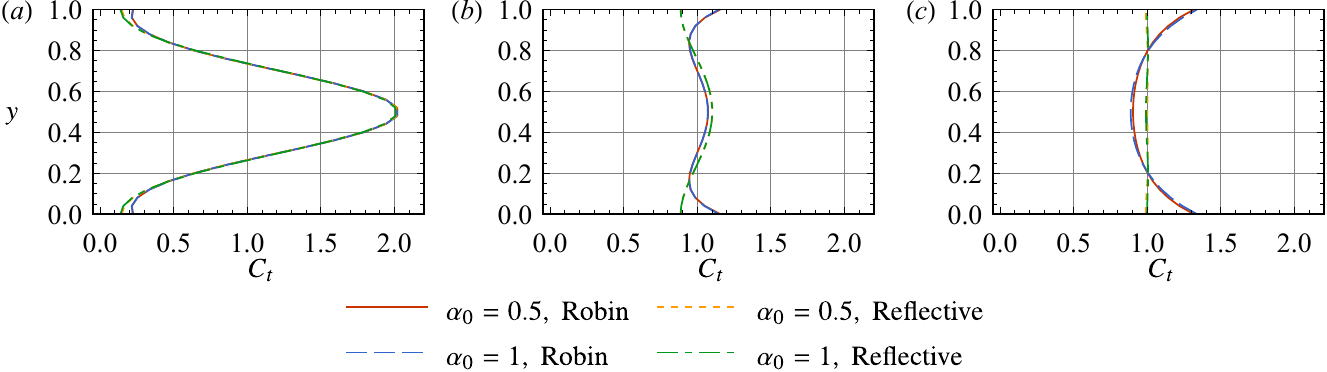}}
	\caption{Transverse distributions $C_t(y,t)$ of ellipsoidal particles under the Robin condition and reflective condition.
		Sample times: (\textit{a}) $t=0.1$; (\textit{b}) $t=0.3$;
		(\textit{c}) $t=0.5$.
		In all cases, $\mathit{Pe}_s = 1$, $\mathit{Pe}_f=2$.
		\label{fig_P0-z_shape}
	}
\end{figure}

The shear-induced alignment of ellipsoidal particles can significantly change the distribution of  $\theta$. 
However, it has a small influence on the vertical concentration distribution.
As shown in \cref{fig_P0-z_shape}, there are only small differences between the vertical distributions with different shape factors under the same boundary condition.
The curves mainly depend on the type of boundary condition for the considered cases with $\mathit{Pe}_f =2$.
The cloud of particles has reached the near-wall region by swimming before the shear-induced alignment exerts its full influence.
% note that even when the shear rate is large, the influece is still not large.

\subsubsection{Dispersion characteristics}

\begin{figure}
	\centering
	{\includegraphics{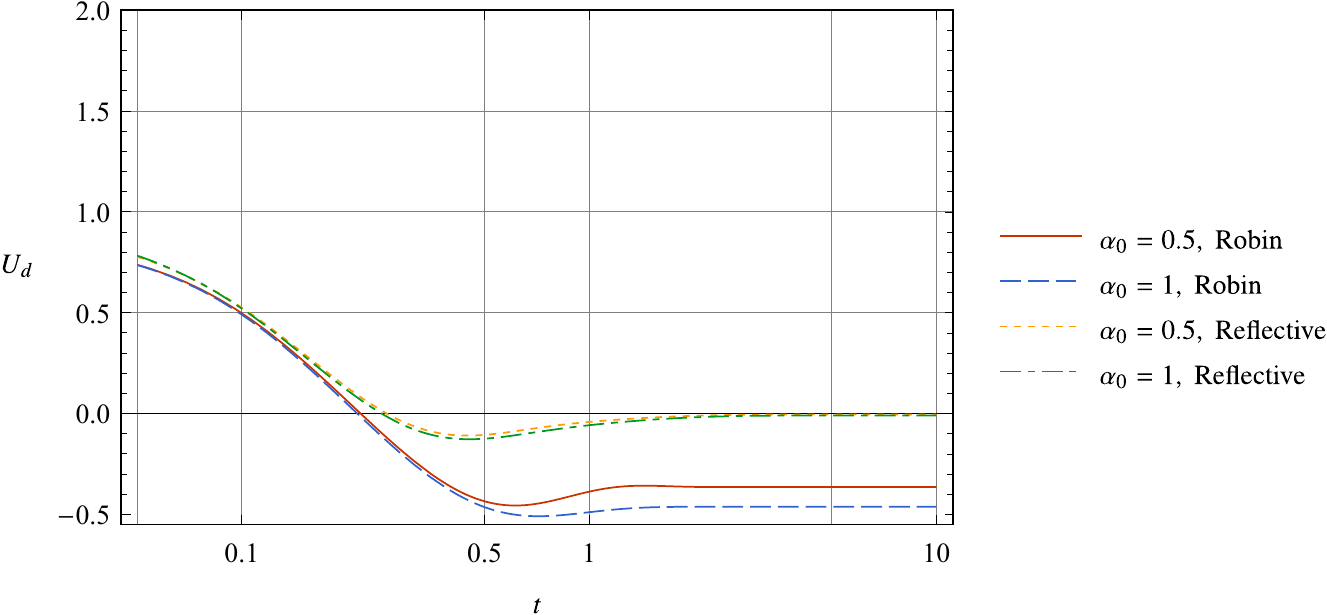}}
	\caption{Temporal evolution of the drift $U_d(t)$ of ellipsoidal particles under 
		the Robin condition and reflective condition.
		In all cases, $\mathit{Pe}_s = 1$, $\mathit{Pe}_f=2$.
		\label{fig_advection_shape}
	}
\end{figure}

We now discuss the temporal evolution of the drift, dispersivity and the skewness for ellipsoidal particles.
First, as shown in \cref{fig_advection_shape}, the effect of the shear-induced alignment on the drift is not large.
In the very initial stage after the point-source release in the middle of the channel, the drift is positive due to the advection effect.
The evolution of the drift of ellipsoidal particles with different shape factors is nearly the same as that of spherical  particles.
At large times, under the reflective condition, the drift of ellipsoidal particles diminishes with time, similar to that of spherical particles.
% why?
Because the vertical distribution is nearly uniform, as shown in \cref{fig_P0-z_shape}, the advection results in a small drift.
The swimming effect is nearly balanced between the preferred directions of the shear-induced alignment.
However, under the Robin conditions, the drift curves deviate from each other at large times.
As discussed in \cref{sec_res_Robin_dispersion}, the wall accumulation leads to a negative drift for spherical particles in the plane Poiseuille flow.
For ellipsoidal particles, the shear-induced alignment further enhances the upstream swimming near the walls.
The stronger the rheotaxis (larger $\alpha_0$), the smaller the drift.

\begin{figure}
	\centering
	{\includegraphics{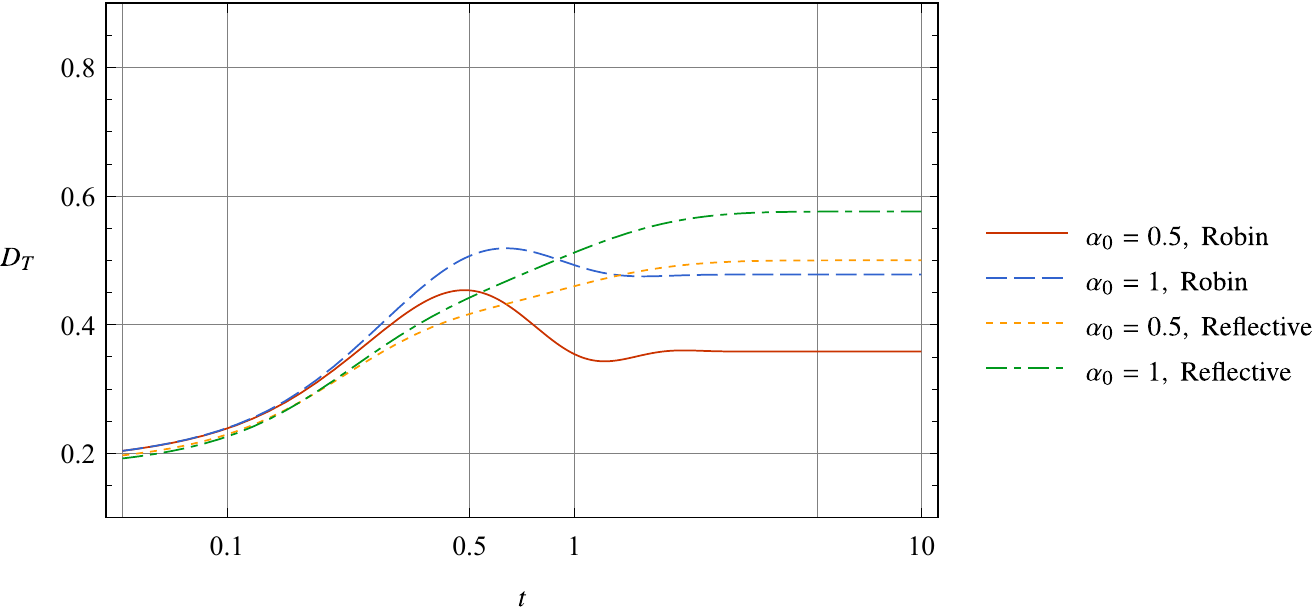}}
	\caption{Temporal evolution of the dispersivity $D_T(t)$ of ellipsoidal particles under the Robin condition and reflective condition.
		In all cases, $\mathit{Pe}_s = 1$, $\mathit{Pe}_f=2$.
		\label{fig_dispersion_shape}
	}
\end{figure}

Next, for the dispersivity, as shown in \cref{fig_dispersion_shape}, the shear-induced alignment can enhance the dispersion process of ellipsoidal particles, especially at large times, for both the reflective boundary condition and the Robin condition.
First, for the reflective condition, the dispersivity increases monotonically with time, similar to the spherical case in \cref{sec_results_shear_dispersion}.
The shear-induced alignment makes the swimming direction of ellipsoidal particles tilt to the streamlines.
Thus the swimming-induce longitudinal dispersivity is larger.
Note that because the shear-induced alignment has a small influence on the vertical concentration distribution, the advection-enhanced dispersivity is almost unaffected by the alignment.
Second, for the Robin condition, the wall accumulation can  suppress the dispersion
process, as discussed in \cref{sec_res_Robin_dispersion} for spherical particles, thus resulting in a drop of the dispersivity as time increases.
The swimming-induce longitudinal dispersivity of  ellipsoidal particles is also enhanced by the alignment, compensating some of the decreases.

\begin{figure}
	\centering
	{\includegraphics{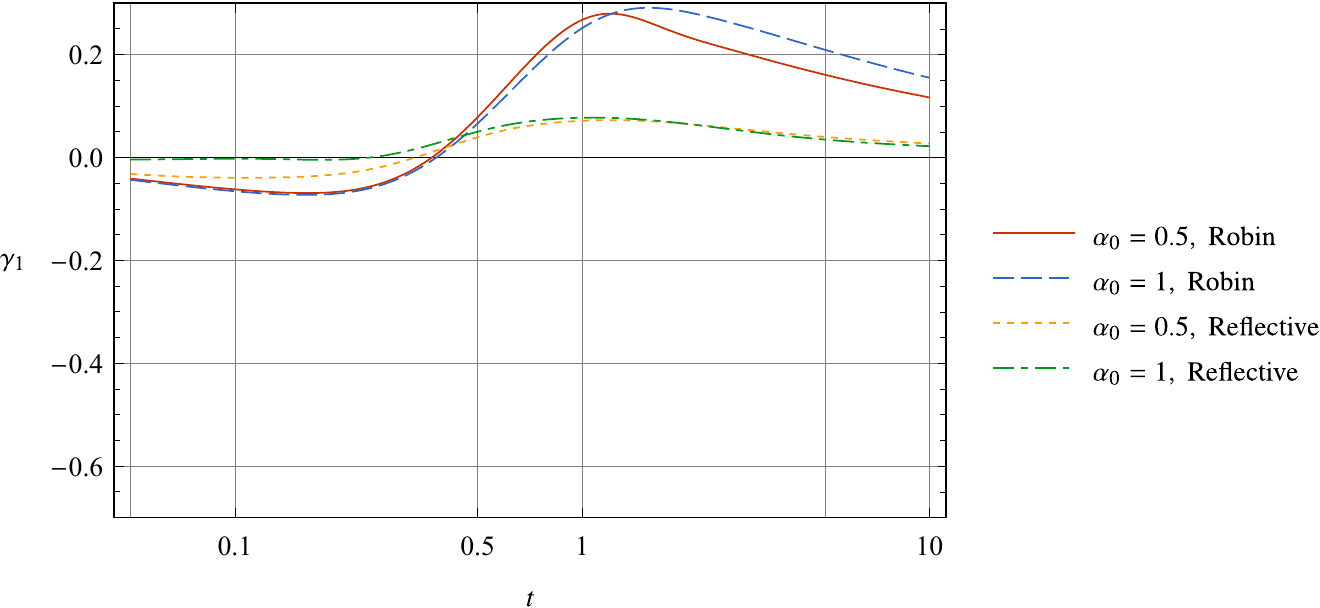}}
	\caption{Temporal evolution of the skewness $\gamma_1(t)$ of ellipsoidal particles under the Robin condition and reflective condition.
		In all cases, $\mathit{Pe}_s = 1$, $\mathit{Pe}_f=2$.
		\label{fig_skew_shape}
	}
\end{figure}

Finally, we discuss the skewness.
Similar to the drift, the shear-induced alignment of ellipsoidal particles has a small influence on the temporal evolution of the skewness, as shown in \cref{fig_skew_shape}.
In the very initial stage, the skewness of ellipsoidal particles is negative due to the advection, the same as that of spherical particles.
Under the reflective condition, the differences between the skewness curves are small, because of the same reasons for the drift:  the advection effects  with different $\alpha_0$ are comparable in nearly uniform vertical distributions, and the swimming effects are nearly balanced between the preferred  directions.
Under the Robin condition,  the reduction  of the swimming-induced diffusion by the strong vorticity-induced rotation near the walls makes the skewness positive, similar to that of spherical particles discussed in \cref{sec_res_Robin_dispersion}.
The shear-induced alignment  ellipsoidal particles can enhance the  swimming-induced diffusion in both the near-wall region and the middle of the channel.
The overall effect enlarges the skewness.
Namely, the cloud of particles swimming downstream disperses faster.

\FloatBarrier 

\section{Conclusions}

% Method
For the transient dispersion process of active particles in confined flows, this work makes the first analytical attempt to investigate the temporal evolution of the dispersion characteristics, including the local distribution in the confined-section--orientation space, the drift, dispersivity and skewness.
To solve the moments of the p.d.f., the classic integral transform method for passive transport problems is not applicable due to the self-propulsion effect.
We introduce the  biorthogonal expansion method to overcome this difficulty.
The auxiliary eigenvalue problem in the local space is solved by the Galerkin method using function series constructed for the reflective boundary condition and the Robin condition for the wall accumulation phenomenon respectively.

% Results
The detailed study on spherical and ellipsoidal swimmers dispersing in a plane Poiseuille flow clearly demonstrates the influences of the swimming, shear flow, wall accumulation and particle shape on the transient dispersion process.
% main findings
% local distribution
After the point-source release at the centreline of the channel,
the local distribution of active particles in the confined-section--orientation space
becomes uniform faster than that of passive particles, as a result of the swimming-induced diffusion.
The vorticity-induced rotation drives spherical particles in the middle of the channel to swim upstream and to perform swing motions.
Under the Robin condition, the wall accumulation is gradually formed as particles spread toward the walls.
If imposing strong shear flow, the accumulation will diminish and the incoming-angle-preferred distribution near the walls will tilt upstream.
The shear-induced alignment of ellipsoidal particles further enhances the upstream-preferred angle distribution near walls but has a less influence on the vertical concentration distribution.

% drift, dispersivity, skewness
For the basic dispersion characteristics, the temporal evolution is complicated under the influences of the swimming, advection and wall accumulation.
Without advection, the drift and the skewness are zero due to the symmetry.
The temporal dispersivity is similar to that in unbounded space, with a anomalous transient stage by the swimming-induced diffusion.
If imposing the plane Poiseuille flow, the advection will lead to a large positive drift and negative skewness in the very initial stage.
The skewness can become positive at large times if the dispersion process is dominant by the swimming-induced diffusion.
For the overall dispersivity, it is not a simple superposition of the shear-enhanced dispersivity and the swimming-induced diffusion.
The wall accumulation can hinder the dispersion process by reducing both the shear-enhanced dispersivity and the swimming-induced diffusion.
However, the accumulation slightly influences the time scale for the Taylor regime.
The shear-induced alignment of ellipsoidal particles can enlarge the dispersivity but have less influences on the drift and the skewness.

% Outlook
It is interesting to extend the current analysis to various situations.
% dilute
For example, this work has only considered very dilute suspensions. 
Future studies on dense suspensions should include particle--particle and particle--fluid interactions.
The temporal evolution of the local distribution plays a key role  in the analyses of the rheological property \citep{takatori_superfluid_2017,saintillan_rheology_2018,nambiar_stress_2019,morris_shear_2020}, self-organization phenomenon,  \citep{vicsek_collective_2012,lushi_fluid_2014,lushi_nonlinear_2018} and hydrodynamic instabilities {\citep{pedley_hydrodynamic_1992,hwang_stability_2014,bees_advances_2020}}.
Besides the self-propulsion effect, taxes of active particles, such as gravitaxis (for gravity), chemotaxis (for chemical gradients) and phototaxis (for light) {\citep{pedley_hydrodynamic_1992,bees_mathematics_2014,goldstein_green_2015}}, probably have great influences on the transient dispersion process.
Moreover, this work only considers a simple type of active particles, whose swimming speed is fixed and the swimming direction undergoes a rotational diffusion process.
The dispersion process of particles with other swimming mechanisms, e.g.\ the run-and-tumble dynamics of \textit{E.\ coli}, is of great interest \citep{berg_random_1993,elgeti_run-and-tumble_2015,vennamneni_shear-induced_2020}.
% or squirmer?
Nevertheless, the influence of particles' swimming behaviours near boundaries is also a fundamental issue.
This work only considers two simple types of boundary condition, the reflective condition and Robin condition, both of which have imposed ideal assumptions. 
In experiments, the observed behaviour at boundaries can be much more complicated \citep{bianchi_holographic_2017,lushi_scattering_2017}, e.g.\ particles sliding along the surface\citep{sipos_hydrodynamic_2015}, scattering off {\citep{volpe_microswimmers_2011,kantsler_ciliary_2013,contino_microalgae_2015}} and the steric repulsion effect {\citep{dehkharghani_bacterial_2019,makarchuk_enhanced_2019}}.
Further work can consider these complex  particle--wall interactions and develop appropriate boundary conditions for the continuum transport model.

%\backsection[Supplementary data]{\label{SupMat}Supplementary material and movies are available at \\https://doi.org/10.1017/jfm.2019...}

%\backsection[Acknowledgements]{Acknowledgements may be included at the end of the paper, before the References section or any appendices. Several anonymous individuals are thanked for contributions to these instructions.}

\backsection[Funding]{This work is supported by the National Natural Science Foundation of China (grant nos 51879002 and 51579004).}

\backsection[Declaration of interests]{The authors report no conflict of interest.}

%\backsection[Data availability statement]{The data that support the findings of this study are openly available in [repository name] at http://doi.org/[doi], reference number [reference number].}

%\backsection[Author ORCID]{Authors may include the ORCID identifers as follows.  F. Smith, https://orcid.org/0000-0001-2345-6789; B. Jones, https://orcid.org/0000-0009-8765-4321}

%\backsection[Author contributions]{Authors may include details of the contributions made by each author to the manuscript, for example, ``A.G. and T.F. derived the theory and T.F. and T.D. performed the simulations.  All authors contributed equally to analysing data and reaching conclusions, and in writing the paper.''}

\appendix

\section{Comparison with Brownian dynamics simulation}
\label{sec_simulations}

To verify the solution of the moments by the biorthogonal expansion method, we perform the Brownian dynamics simulation, which is widely used in numerical studies \citep{croze_dispersion_2013,chilukuri_dispersion_2015,apaza_ballistic_2016}.
Attention should be paid to the treatment of boundary conditions for the reflective and Robin types.

According to the dimensionless governing equation \cref{eq probability conservation simple}, the corresponding stochastic differential equations for the coordinates of a swimmer $(x(t), y(t), \theta(t))$ are
\begin{equation}
	\left\{\begin{array}{l}
		\dfrac{\mathrm{d} x}{\mathrm{d} t} = \mathit{Pe}_f u (y) + \mathit{Pe}_s
		\cos \theta + \sqrt{2 D_t}  \dfrac{\mathrm{d} W_x}{\mathrm{d} t},\\
		\dfrac{\mathrm{d} y}{\mathrm{d} t} = \mathit{Pe}_s \sin \theta + \sqrt{2
			D_t}  \dfrac{\mathrm{d} W_y}{\mathrm{d} t},\\
		\dfrac{\mathrm{d} \theta}{\mathrm{d} t} = \Omega (y, \theta) + \sqrt{2
			D_r}  \dfrac{\mathrm{d} W_{\theta}}{\mathrm{d} t},
	\end{array}\right.
\end{equation}
where $W_x(t)$, $W_y(t)$ and $W_{\theta}(t)$ are independent standard Brownian motions.
We simply apply a forward Euler scheme with time step $\Delta t$ for discretization.
The $n$-th step coordinates of the swimmer are denoted as $(x_n, y_n, \theta_n)$.
For the typical reflective boundary condition \cref{eq_reflective_BC}, if the swimmer exceeds the boundaries, then
\begin{equation}
	\left\{\begin{array}{ll}
		y_n \rightarrow 2 - y_n, \quad \theta_n \rightarrow - \theta_n, &
		\mathrm{for} \; y_n > 1,\\
		y_n \rightarrow - y_n, \quad \theta_n \rightarrow - \theta_n, &
		\mathrm{for} \; y_n <0.
	\end{array}\right.
\end{equation}
where $\rightarrow$ means assignment.This treatment is common in the Brownian dynamics simulations \citep{volpe_simulation_2014,bechinger_active_2016}.
For the Robin condition to account for the wall accumulation of swimmers, 
\begin{equation}
	\left\{\begin{array}{ll}
		y_n \rightarrow 2 - y_n, \quad \theta_n \rightarrow \theta_n, &
		\mathrm{for} \; y_n > 1,\\
		y_n \rightarrow - y_n, \quad \theta_n \rightarrow \theta_n, & \mathrm{for}
		\; y_n > 1.
	\end{array}\right.
\end{equation}
Note that the swimming direction is not reversed as the reflective condition after the collision with a wall.
In fact, the Robin condition, or the no-penetration condition for the probability flux, can be viewed as a ``reflective'' condition for the boundaries in the phase space with $\theta$ treated as an extra position coordinate.
Another similar treatment puts the swimmer that exceeds a wall back exactly at the wall \citep{duzgun_active_2018,peng_upstream_2020}.

\begin{figure}
	\centering
	{\includegraphics{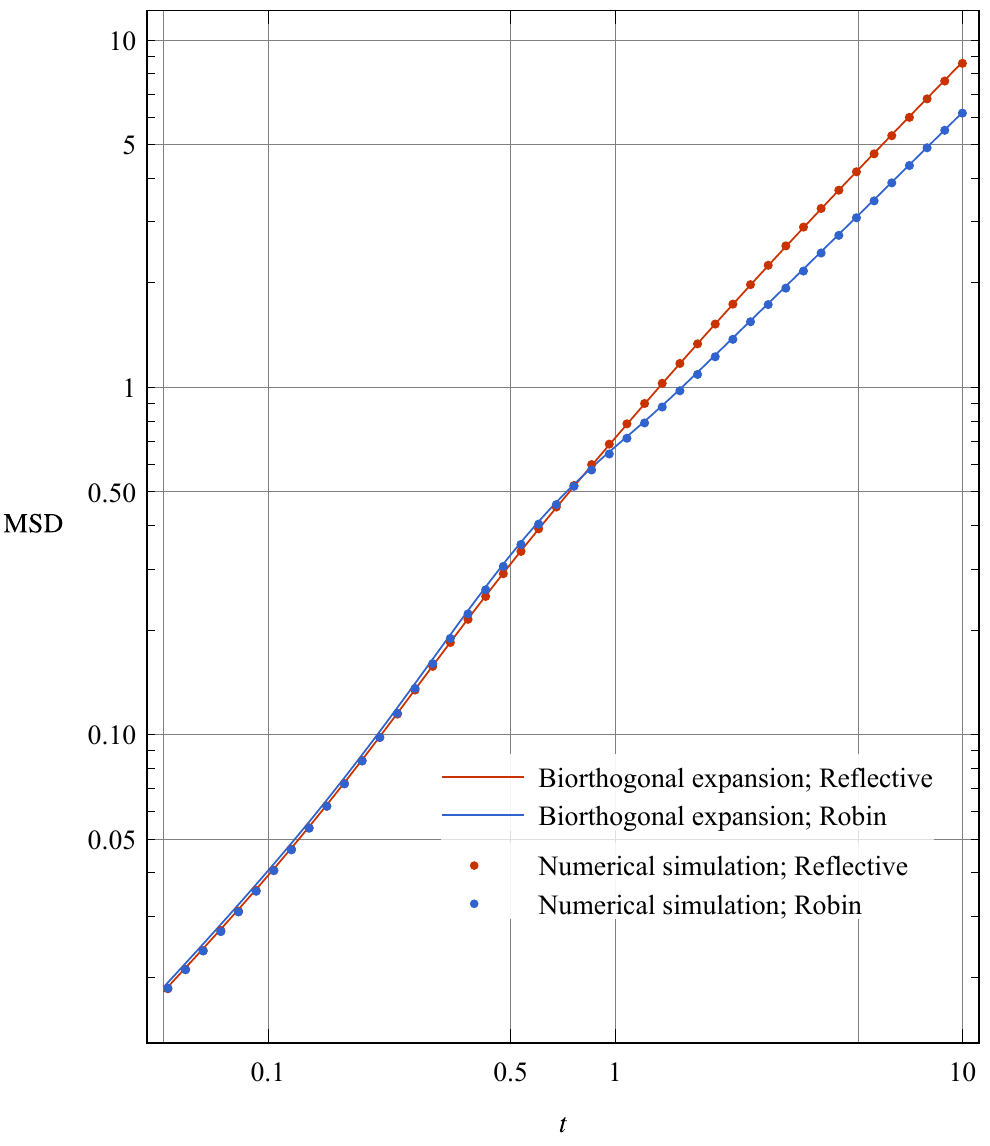}}
	\caption{Comparisons between the temporal evolution of $\mathrm{MSD}$ of spherical particles by the biorthogonal expansion and 
		Brownian dynamics simulation under different boundary conditions.
		`Robin' denotes the Robin condition, and `Reflective' denotes the reflective condition.
		In all cases, $\mathit{Pe}_s=1$, $\mathit{Pe}_f=2$.
		\label{fig_compare_BC}
	}
\end{figure}

In the simulation, we use a small time step $\Delta t = 10^{-3}$ to capture the transient transport process.
Swimmer are initially put at $y_1=\frac{1}{2}$ with $\theta_0$ uniformly distributed in the interval $[-\upi, \upi)$.
$10^{5}$ trajectories are simulated for each case.
As shown in \cref{fig_compare_BC},  the analytical results of the MSD of spherical swimmers by the biorthogonal expansion is in agreement with the numerical results by the Brownian dynamics simulations under both the two types of boundary condition.

\FloatBarrier
\bibliographystyle{jfm} % Note the spaces between the initials
\bibliography{E:/Research/More/Zotero/mylibraryAbbreviated}

\begin{thebibliography}{112}
\expandafter\ifx\csname natexlab\endcsname\relax\def\natexlab#1{#1}\fi
\def\au#1{#1} \def\ed#1{#1} \def\yr#1{#1}\def\at#1{#1}\def\jt#1{\textit{#1}}
  \def\bt#1{#1}\def\bvol#1{\textbf{#1}} \def\vol#1{#1} \def\pg#1{#1}
  \def\publ#1{#1}\def\arxiv#1{#1}\def\org#1{#1}\def\st#1{\textit{#1}}

\bibitem[Aci{\'e}n {\em et~al.\/}(2017)Aci{\'e}n, Molina, Reis, Torzillo,
  Zittelli, Sep{\'u}lveda \& Masoj{\'i}dek]{acien_photobioreactors_2017}
{\sc \au{Aci{\'e}n, F.~G.}, \au{Molina, E.}, \au{Reis, A.}, \au{Torzillo, G.},
  \au{Zittelli, G.~C.}, \au{Sep{\'u}lveda, C.} \& \au{Masoj{\'i}dek, J.}}
  \yr{2017}  \at{Photobioreactors for the production of microalgae}.  \bt{In
  {\em Microalgae-{{Based Biofuels}} and {{Bioproducts}}\/} (ed.
  \ed{C.~{Gonzalez-Fernandez} \& R.~Mu{\~n}oz})},  \pg{pp. 1--44}.
  \publ{{Duxford}: {Woodhead Publishing}}.

\bibitem[{Alonso-Matilla} {\em et~al.\/}(2019){Alonso-Matilla}, Chakrabarti \&
  Saintillan]{alonso-matilla_transport_2019}
{\sc \au{{Alonso-Matilla}, R.}, \au{Chakrabarti, B.} \& \au{Saintillan, D.}}
  \yr{2019}  \at{Transport and dispersion of active particles in periodic
  porous media}.  \jt{Phys. Rev. Fluids}  \bvol{4}~(4),  \pg{043101}.

\bibitem[Aminian {\em et~al.\/}(2016)Aminian, Bernardi, Camassa, Harris \&
  McLaughlin]{aminian_how_2016}
{\sc \au{Aminian, M.}, \au{Bernardi, F.}, \au{Camassa, R.}, \au{Harris, D.~M.}
  \& \au{McLaughlin, R.~M.}} \yr{2016}  \at{How boundaries shape chemical
  delivery in microfluidics}.  \jt{Science}  \bvol{354}~(6317),
  \pg{1252--1256}.

\bibitem[Ao {\em et~al.\/}(2014)Ao, Ghosh, Li, Schmid, H{\"a}nggi \&
  Marchesoni]{ao_active_2014}
{\sc \au{Ao, X.}, \au{Ghosh, P.}, \au{Li, Y.}, \au{Schmid, G.}, \au{H{\"a}nggi,
  P.} \& \au{Marchesoni, F.}} \yr{2014}  \at{Active {{Brownian}} motion in a
  narrow channel}.  \jt{Eur. Phys. J. Spec. Top.}  \bvol{223}~(14),
  \pg{3227--3242}.

\bibitem[Apaza \& Sandoval(2016)]{apaza_ballistic_2016}
{\sc \au{Apaza, L.} \& \au{Sandoval, M.}} \yr{2016}  \at{Ballistic behavior and
  trapping of self-driven particles in a {{Poiseuille}} flow}.  \jt{Phys. Rev.
  E}  \bvol{93}~(6),  \pg{062602}.

\bibitem[Aris(1956)]{aris_dispersion_1956}
{\sc \au{Aris, R.}} \yr{1956}  \at{On the dispersion of a solute in a fluid
  flowing through a tube}.  \jt{Proc. R. Soc. Lond. Math. Phys. Eng. Sci.}
  \bvol{235}~(1200),  \pg{67--77}.

\bibitem[Barton(1983)]{barton_method_1983}
{\sc \au{Barton, N.~G.}} \yr{1983}  \at{On the method of moments for solute
  dispersion}.  \jt{J. Fluid Mech.}  \bvol{126},  \pg{205--218}.

\bibitem[Bearon {\em et~al.\/}(2012)Bearon, Bees \& Croze]{bearon_biased_2012}
{\sc \au{Bearon, R.~N.}, \au{Bees, M.~A.} \& \au{Croze, O.~A.}} \yr{2012}
  \at{Biased swimming cells do not disperse in pipes as tracers: a population
  model based on microscale behaviour}.  \jt{Phys. Fluids}  \bvol{24}~(12),
  \pg{121902}.

\bibitem[Bearon \& Hazel(2015)]{bearon_trapping_2015}
{\sc \au{Bearon, R.~N.} \& \au{Hazel, A.~L.}} \yr{2015}  \at{The trapping in
  high-shear regions of slender bacteria undergoing chemotaxis in a channel}.
  \jt{J. Fluid Mech.}  \bvol{771},  \pg{R3}.

\bibitem[Bearon {\em et~al.\/}(2011)Bearon, Hazel \&
  Thorn]{bearon_spatial_2011}
{\sc \au{Bearon, R.~N.}, \au{Hazel, A.~L.} \& \au{Thorn, G.~J.}} \yr{2011}
  \at{The spatial distribution of gyrotactic swimming micro-organisms in
  laminar flow fields}.  \jt{J. Fluid Mech.}  \bvol{680},  \pg{602--635}.

\bibitem[Bechinger {\em et~al.\/}(2016)Bechinger, Di~Leonardo, L{\"o}wen,
  Reichhardt, Volpe \& Volpe]{bechinger_active_2016}
{\sc \au{Bechinger, C.}, \au{Di~Leonardo, R.}, \au{L{\"o}wen, H.},
  \au{Reichhardt, C.}, \au{Volpe, G.} \& \au{Volpe, G.}} \yr{2016}  \at{Active
  particles in complex and crowded environments}.  \jt{Rev. Mod. Phys.}
  \bvol{88}~(4),  \pg{045006}.

\bibitem[Bees(2020)]{bees_advances_2020}
{\sc \au{Bees, M.~A.}} \yr{2020}  \at{Advances in bioconvection}.  \jt{Annu.
  Rev. Fluid Mech.}  \bvol{52}~(1),  \pg{449--476}.

\bibitem[Bees \& Croze(2010)]{bees_dispersion_2010}
{\sc \au{Bees, M.~A.} \& \au{Croze, O.~A.}} \yr{2010}  \at{Dispersion of biased
  swimming micro-organisms in a fluid flowing through a tube}.  \jt{Proc. R.
  Soc. Lond. Math. Phys. Eng. Sci.}  \bvol{466}~(2119),  \pg{2057--2077}.

\bibitem[Bees \& Croze(2014)]{bees_mathematics_2014}
{\sc \au{Bees, M.~A.} \& \au{Croze, O.~A.}} \yr{2014}  \at{Mathematics for
  streamlined biofuel production from unicellular algae}.  \jt{Biofuels}
  \bvol{5}~(1),  \pg{53--65}.

\bibitem[Berg(1993)]{berg_random_1993}
{\sc \au{Berg, H.~C.}} \yr{1993} {\em Random {{Walks}} in {{Biology}}\/},
  revised edn.  \publ{{Princeton}: {Princeton University Press}}.

\bibitem[Berg \& Turner(1990)]{berg_chemotaxis_1990}
{\sc \au{Berg, H.~C.} \& \au{Turner, L.}} \yr{1990}  \at{Chemotaxis of bacteria
  in glass capillary arrays. {{{\emph{Escherichia}}}}{\emph{ coli}}, motility,
  microchannel plate, and light scattering}.  \jt{Biophys. J.}  \bvol{58}~(4),
  \pg{919--930}.

\bibitem[Berke {\em et~al.\/}(2008)Berke, Turner, Berg \&
  Lauga]{berke_hydrodynamic_2008}
{\sc \au{Berke, A.~P.}, \au{Turner, L.}, \au{Berg, H.~C.} \& \au{Lauga, E.}}
  \yr{2008}  \at{Hydrodynamic attraction of swimming microorganisms by
  surfaces}.  \jt{Phys. Rev. Lett.}  \bvol{101}~(3),  \pg{038102}.

\bibitem[Berlyand {\em et~al.\/}(2020)Berlyand, Jabin, Potomkin \&
  Ratajczyk]{berlyand_kinetic_2020}
{\sc \au{Berlyand, L.}, \au{Jabin, P.-E.}, \au{Potomkin, M.} \& \au{Ratajczyk,
  E.}} \yr{2020}  \at{A kinetic approach to active rods dynamics in confined
  domains}.  \jt{Multiscale Model. Simul.}  \bvol{18}~(1),  \pg{1--20}.

\bibitem[Bianchi {\em et~al.\/}(2017)Bianchi, Saglimbeni \&
  Di~Leonardo]{bianchi_holographic_2017}
{\sc \au{Bianchi, S.}, \au{Saglimbeni, F.} \& \au{Di~Leonardo, R.}} \yr{2017}
  \at{Holographic imaging reveals the mechanism of wall entrapment in swimming
  bacteria}.  \jt{Phys. Rev. X}  \bvol{7}~(1),  \pg{011010}.

\bibitem[Brenner(1982)]{brenner_general_1982a}
{\sc \au{Brenner, H.}} \yr{1982}  \at{A general theory of {{Taylor}} dispersion
  phenomena. {{II}}. {{An}} extension}.  \jt{Physico-Chem. Hydrodyn.}
  \bvol{3}~(2),  \pg{139--157}.

\bibitem[Brenner \& Edwards(1993)]{brenner_macrotransport_1993}
{\sc \au{Brenner, H.} \& \au{Edwards, D.~A.}} \yr{1993} {\em Macrotransport
  {{Processes}}\/}.  \publ{{Stoneham}: {Butterworth-Heinemann}}.

\bibitem[Brezinski(1991)]{brezinski_biorthogonality_1991}
{\sc \au{Brezinski, C.}} \yr{1991} {\em Biorthogonality and its
  {{Applications}} to {{Numerical Analysis}}\/}.  \publ{{New York}: {Marcel
  Dekker}}.

\bibitem[Brosseau {\em et~al.\/}(2019)Brosseau, Usabiaga, Lushi, Wu, Ristroph,
  Zhang, Ward \& Shelley]{brosseau_relating_2019}
{\sc \au{Brosseau, Q.}, \au{Usabiaga, F.~B.}, \au{Lushi, E.}, \au{Wu, Y.},
  \au{Ristroph, L.}, \au{Zhang, J.}, \au{Ward, M.} \& \au{Shelley, M.~J.}}
  \yr{2019}  \at{Relating rheotaxis and hydrodynamic actuation using asymmetric
  gold-platinum phoretic rods}.  \jt{Phys. Rev. Lett.}  \bvol{123}~(17),
  \pg{178004}.

\bibitem[Camassa {\em et~al.\/}(2010)Camassa, Lin \&
  McLaughlin]{camassa_exact_2010}
{\sc \au{Camassa, R.}, \au{Lin, Z.} \& \au{McLaughlin, R.~M.}} \yr{2010}
  \at{The exact evolution of the scalar variance in pipe and channel flow}.
  \jt{Commun. Math. Sci.}  \bvol{8}~(2),  \pg{601--626}.

\bibitem[Chatwin(1970)]{chatwin_approach_1970}
{\sc \au{Chatwin, P.~C.}} \yr{1970}  \at{The approach to normality of the
  concentration distribution of a solute in a solvent flowing along a straight
  pipe}.  \jt{J. Fluid Mech.}  \bvol{43}~(2),  \pg{321--352}.

\bibitem[Chilukuri {\em et~al.\/}(2014)Chilukuri, Collins \&
  Underhill]{chilukuri_impact_2014}
{\sc \au{Chilukuri, S.}, \au{Collins, C.~H.} \& \au{Underhill, P.~T.}}
  \yr{2014}  \at{Impact of external flow on the dynamics of swimming
  microorganisms near surfaces}.  \jt{J. Phys. Condens. Matter}
  \bvol{26}~(11),  \pg{115101}.

\bibitem[Chilukuri {\em et~al.\/}(2015)Chilukuri, Collins \&
  Underhill]{chilukuri_dispersion_2015}
{\sc \au{Chilukuri, S.}, \au{Collins, C.~H.} \& \au{Underhill, P.~T.}}
  \yr{2015}  \at{Dispersion of flagellated swimming microorganisms in planar
  {{Poiseuille}} flow}.  \jt{Phys. Fluids}  \bvol{27}~(3),  \pg{031902}.

\bibitem[Chisti(2007)]{chisti_biodiesel_2007}
{\sc \au{Chisti, Y.}} \yr{2007}  \at{Biodiesel from microalgae}.
  \jt{Biotechnol. Adv.}  \bvol{25}~(3),  \pg{294--306}.

\bibitem[Contino {\em et~al.\/}(2015)Contino, Lushi, Tuval, Kantsler \&
  Polin]{contino_microalgae_2015}
{\sc \au{Contino, M.}, \au{Lushi, E.}, \au{Tuval, I.}, \au{Kantsler, V.} \&
  \au{Polin, M.}} \yr{2015}  \at{Microalgae scatter off solid surfaces by
  hydrodynamic and contact forces}.  \jt{Phys. Rev. Lett.}  \bvol{115}~(25),
  \pg{258102}.

\bibitem[Costanzo {\em et~al.\/}(2012)Costanzo, Di~Leonardo, Ruocco \&
  Angelani]{costanzo_transport_2012}
{\sc \au{Costanzo, A.}, \au{Di~Leonardo, R.}, \au{Ruocco, G.} \& \au{Angelani,
  L.}} \yr{2012}  \at{Transport of self-propelling bacteria in micro-channel
  flow}.  \jt{J. Phys. Condens. Matter}  \bvol{24}~(6),  \pg{065101}.

\bibitem[Croze {\em et~al.\/}(2017)Croze, Bearon \&
  Bees]{croze_gyrotactic_2017}
{\sc \au{Croze, O.~A.}, \au{Bearon, R.~N.} \& \au{Bees, M.~A.}} \yr{2017}
  \at{Gyrotactic swimmer dispersion in pipe flow: testing the theory}.  \jt{J.
  Fluid Mech.}  \bvol{816},  \pg{481--506}.

\bibitem[Croze {\em et~al.\/}(2013)Croze, Sardina, Ahmed, Bees \&
  Brandt]{croze_dispersion_2013}
{\sc \au{Croze, O.~A.}, \au{Sardina, G.}, \au{Ahmed, M.}, \au{Bees, M.~A.} \&
  \au{Brandt, L.}} \yr{2013}  \at{Dispersion of swimming algae in laminar and
  turbulent channel flows: consequences for photobioreactors}.  \jt{J. R. Soc.
  Interface}  \bvol{10}~(81),  \pg{20121041}.

\bibitem[Dehkharghani {\em et~al.\/}(2019)Dehkharghani, Waisbord, Dunkel \&
  Guasto]{dehkharghani_bacterial_2019}
{\sc \au{Dehkharghani, A.}, \au{Waisbord, N.}, \au{Dunkel, J.} \& \au{Guasto,
  J.~S.}} \yr{2019}  \at{Bacterial scattering in microfluidic crystal flows
  reveals giant active {{Taylor}}\textendash{{Aris}} dispersion}.  \jt{PNAS}
  \bvol{116}~(23),  \pg{11119--11124}.

\bibitem[Doi \& Edwards(1988)]{doi_brownian_1988}
{\sc \au{Doi, M.} \& \au{Edwards, S.~F.}} \yr{1988}  \at{Brownian motion}.
  \bt{In {\em The {{Theory}} of {{Polymer Dynamics}}\/}},  \pg{pp. 46--90}.
  \publ{{Oxford}: {Oxford University Press}}.

\bibitem[Drescher {\em et~al.\/}(2011)Drescher, Dunkel, Cisneros, Ganguly \&
  Goldstein]{drescher_fluid_2011}
{\sc \au{Drescher, K.}, \au{Dunkel, J.}, \au{Cisneros, L.~H.}, \au{Ganguly, S.}
  \& \au{Goldstein, R.~E.}} \yr{2011}  \at{Fluid dynamics and noise in
  bacterial cell\textendash cell and cell\textendash surface scattering}.
  \jt{PNAS}  \bvol{108}~(27),  \pg{10940--10945}.

\bibitem[Durham \& Stocker(2012)]{durham_thin_2012}
{\sc \au{Durham, W.~M.} \& \au{Stocker, R.}} \yr{2012}  \at{Thin phytoplankton
  layers: characteristics, mechanisms, and consequences}.  \jt{Annu. Rev. Mar.
  Sci.}  \bvol{4}~(1),  \pg{177--207}.

\bibitem[Duzgun \& Selinger(2018)]{duzgun_active_2018}
{\sc \au{Duzgun, A.} \& \au{Selinger, J.~V.}} \yr{2018}  \at{Active
  {{Brownian}} particles near straight or curved walls: pressure and boundary
  layers}.  \jt{Phys. Rev. E}  \bvol{97}~(3),  \pg{032606}.

\bibitem[Elgeti \& Gompper(2013)]{elgeti_wall_2013}
{\sc \au{Elgeti, J.} \& \au{Gompper, G.}} \yr{2013}  \at{Wall accumulation of
  self-propelled spheres}.  \jt{EPL}  \bvol{101}~(4),  \pg{48003}.

\bibitem[Elgeti \& Gompper(2015)]{elgeti_run-and-tumble_2015}
{\sc \au{Elgeti, J.} \& \au{Gompper, G.}} \yr{2015}  \at{Run-and-tumble
  dynamics of self-propelled particles in confinement}.  \jt{EPL}
  \bvol{109}~(5),  \pg{58003}.

\bibitem[Enculescu \& Stark(2011)]{enculescu_active_2011}
{\sc \au{Enculescu, M.} \& \au{Stark, H.}} \yr{2011}  \at{Active colloidal
  suspensions exhibit polar order under gravity}.  \jt{Phys. Rev. Lett.}
  \bvol{107}~(5),  \pg{058301}.

\bibitem[Ezhilan \& Saintillan(2015)]{ezhilan_transport_2015}
{\sc \au{Ezhilan, B.} \& \au{Saintillan, D.}} \yr{2015}  \at{Transport of a
  dilute active suspension in pressure-driven channel flow}.  \jt{J. Fluid
  Mech.}  \bvol{777},  \pg{482--522}.

\bibitem[Foister \& {van de Ven}(1980)]{foister_diffusion_1980}
{\sc \au{Foister, R.~T.} \& \au{{van de Ven}, T. G.~M.}} \yr{1980}
  \at{Diffusion of {{Brownian}} particles in shear flows}.  \jt{J. Fluid Mech.}
   \bvol{96}~(1),  \pg{105--132}.

\bibitem[Frankel \& Brenner(1989)]{frankel_foundations_1989}
{\sc \au{Frankel, I.} \& \au{Brenner, H.}} \yr{1989}  \at{On the foundations of
  generalized {{Taylor}} dispersion theory}.  \jt{J. Fluid Mech.}  \bvol{204},
  \pg{97--119}.

\bibitem[Fung {\em et~al.\/}(2020)Fung, Bearon \& Hwang]{fung_bifurcation_2020}
{\sc \au{Fung, L.}, \au{Bearon, R.~N.} \& \au{Hwang, Y.}} \yr{2020}
  \at{Bifurcation and stability of downflowing gyrotactic micro-organism
  suspensions in a vertical pipe}.  \jt{J. Fluid Mech.}  \bvol{902},  \pg{A26}.

\bibitem[Ghosh {\em et~al.\/}(2013)Ghosh, Misko, Marchesoni \&
  Nori]{ghosh_self-propelled_2013}
{\sc \au{Ghosh, P.~K.}, \au{Misko, V.~R.}, \au{Marchesoni, F.} \& \au{Nori,
  F.}} \yr{2013}  \at{Self-propelled {{Janus}} particles in a ratchet:
  numerical simulations}.  \jt{Phys. Rev. Lett.}  \bvol{110}~(26),
  \pg{268301}.

\bibitem[Gill(1967)]{gill_note_1967}
{\sc \au{Gill, W.~N.}} \yr{1967}  \at{A note on the solution of transient
  dispersion problems}.  \jt{Proc. R. Soc. Lond. Math. Phys. Eng. Sci.}
  \bvol{298}~(1454),  \pg{335--339}.

\bibitem[Gill \& Sankarasubramanian(1970)]{gill_exact_1970}
{\sc \au{Gill, W.~N.} \& \au{Sankarasubramanian, R.}} \yr{1970}  \at{Exact
  analysis of unsteady convective diffusion}.  \jt{Proc. R. Soc. Lond. Math.
  Phys. Eng. Sci.}  \bvol{316}~(1526),  \pg{341--350}.

\bibitem[Goldstein(2015)]{goldstein_green_2015}
{\sc \au{Goldstein, R.~E.}} \yr{2015}  \at{Green algae as model organisms for
  biological fluid dynamics}.  \jt{Annu. Rev. Fluid Mech.}  \bvol{47}~(1),
  \pg{343--375}.

\bibitem[Guazzelli \& Morris(2012)]{guazzelli_physical_2012}
{\sc \au{Guazzelli, {\'E}.} \& \au{Morris, J.~F.}} \yr{2012} {\em A {{Physical
  Introduction}} to {{Suspension Dynamics}}\/}.  \publ{{Cambridge}: {Cambridge
  University Press}}.

\bibitem[Hill \& Bees(2002)]{hill_taylor_2002}
{\sc \au{Hill, N.~A.} \& \au{Bees, M.~A.}} \yr{2002}  \at{Taylor dispersion of
  gyrotactic swimming micro-organisms in a linear flow}.  \jt{Phys. Fluids}
  \bvol{14}~(8),  \pg{2598--2605}.

\bibitem[Hill \& Pedley(2005)]{hill_bioconvection_2005}
{\sc \au{Hill, N.~A.} \& \au{Pedley, T.~J.}} \yr{2005}  \at{Bioconvection}.
  \jt{Fluid Dyn. Res.}  \bvol{37}~(1-2),  \pg{1--20}.

\bibitem[Howse {\em et~al.\/}(2007)Howse, Jones, Ryan, Gough, Vafabakhsh \&
  Golestanian]{howse_self-motile_2007}
{\sc \au{Howse, J.~R.}, \au{Jones, R. A.~L.}, \au{Ryan, A.~J.}, \au{Gough, T.},
  \au{Vafabakhsh, R.} \& \au{Golestanian, R.}} \yr{2007}  \at{Self-motile
  colloidal particles: from directed propulsion to random walk}.  \jt{Phys.
  Rev. Lett.}  \bvol{99}~(4),  \pg{048102}.

\bibitem[Hwang \& Pedley(2014)]{hwang_stability_2014}
{\sc \au{Hwang, Y.} \& \au{Pedley, T.~J.}} \yr{2014}  \at{Stability of
  downflowing gyrotactic microorganism suspensions in a two-dimensional
  vertical channel}.  \jt{J. Fluid Mech.}  \bvol{749},  \pg{750--777}.

\bibitem[Jeffery(1922)]{jeffery_motion_1922}
{\sc \au{Jeffery, G.~B.}} \yr{1922}  \at{The motion of ellipsoidal particles
  immersed in a viscous fluid}.  \jt{Proc. R. Soc. Lond. Math. Phys. Eng. Sci.}
   \bvol{102}~(715),  \pg{161--179}.

\bibitem[Jiang \& Chen(2019{\natexlab{{\em a\/}}})]{jiang_dispersion_2019}
{\sc \au{Jiang, W.} \& \au{Chen, G.}} \yr{2019{\natexlab{{\em a\/}}}}
  \at{Dispersion of active particles in confined unidirectional flows}.  \jt{J.
  Fluid Mech.}  \bvol{877},  \pg{1--34}.

\bibitem[Jiang \& Chen(2019{\natexlab{{\em b\/}}})]{jiang_solute_2019}
{\sc \au{Jiang, W.} \& \au{Chen, G.}} \yr{2019{\natexlab{{\em b\/}}}}
  \at{Solute transport in two-zone packed tube flow: long-time asymptotic
  expansion}.  \jt{Phys. Fluids}  \bvol{31}~(4),  \pg{043303}.

\bibitem[Jiang \& Chen(2020)]{jiang_dispersion_2020}
{\sc \au{Jiang, W.} \& \au{Chen, G.}} \yr{2020}  \at{Dispersion of gyrotactic
  micro-organisms in pipe flows}.  \jt{J. Fluid Mech.}  \bvol{889},  \pg{A18}.

\bibitem[Kantsler {\em et~al.\/}(2013)Kantsler, Dunkel, Polin \&
  Goldstein]{kantsler_ciliary_2013}
{\sc \au{Kantsler, V.}, \au{Dunkel, J.}, \au{Polin, M.} \& \au{Goldstein,
  R.~E.}} \yr{2013}  \at{Ciliary contact interactions dominate surface
  scattering of swimming eukaryotes}.  \jt{PNAS}  \bvol{110}~(4),
  \pg{1187--1192}.

\bibitem[Latini \& Bernoff(2001)]{latini_transient_2001}
{\sc \au{Latini, M.} \& \au{Bernoff, A.~J.}} \yr{2001}  \at{Transient anomalous
  diffusion in {{Poiseuille}} flow}.  \jt{J. Fluid Mech.}  \bvol{441},
  \pg{399--411}.

\bibitem[Lauga {\em et~al.\/}(2006)Lauga, DiLuzio, Whitesides \&
  Stone]{lauga_swimming_2006}
{\sc \au{Lauga, E.}, \au{DiLuzio, W.~R.}, \au{Whitesides, G.~M.} \& \au{Stone,
  H.~A.}} \yr{2006}  \at{Swimming in circles: motion of bacteria near solid
  boundaries}.  \jt{Biophys. J.}  \bvol{90}~(2),  \pg{400--412}.

\bibitem[Lauga \& Powers(2009)]{lauga_hydrodynamics_2009}
{\sc \au{Lauga, E.} \& \au{Powers, T.~R.}} \yr{2009}  \at{The hydrodynamics of
  swimming microorganisms}.  \jt{Rep. Prog. Phys.}  \bvol{72}~(9),
  \pg{096601}.

\bibitem[Leal \& Hinch(1972)]{leal_rheology_1972}
{\sc \au{Leal, L.~G.} \& \au{Hinch, E.~J.}} \yr{1972}  \at{The rheology of a
  suspension of nearly spherical particles subject to {{Brownian}} rotations}.
  \jt{J. Fluid Mech.}  \bvol{55}~(4),  \pg{745--765}.

\bibitem[Li {\em et~al.\/}(2008)Li, Tam \& Tang]{li_amplified_2008}
{\sc \au{Li, G.}, \au{Tam, L.-K.} \& \au{Tang, J.~X.}} \yr{2008}  \at{Amplified
  effect of {{Brownian}} motion in bacterial near-surface swimming}.  \jt{PNAS}
   \bvol{105}~(47),  \pg{18355--18359}.

\bibitem[Li \& Tang(2009)]{li_accumulation_2009}
{\sc \au{Li, G.} \& \au{Tang, J.~X.}} \yr{2009}  \at{Accumulation of
  microswimmers near a surface mediated by collision and rotational
  {{Brownian}} motion}.  \jt{Phys. Rev. Lett.}  \bvol{103}~(7),  \pg{078101}.

\bibitem[Lighthill(1966)]{lighthill_initial_1966}
{\sc \au{Lighthill, M.~J.}} \yr{1966}  \at{Initial development of diffusion in
  {{Poiseuille}} flow}.  \jt{IMA J. Appl. Math.}  \bvol{2}~(1),  \pg{97--108}.

\bibitem[Liu {\em et~al.\/}(2012)Liu, Liu, Johnson, Yi \&
  Huang]{liu_effects_2012}
{\sc \au{Liu, L.}, \au{Liu, D.}, \au{Johnson, D.~M.}, \au{Yi, Z.} \& \au{Huang,
  Y.}} \yr{2012}  \at{Effects of vertical mixing on phytoplankton blooms in
  {{Xiangxi Bay}} of {{Three Gorges Reservoir}}: implications for management}.
  \jt{Water Res.}  \bvol{46}~(7),  \pg{2121--2130}.

\bibitem[Lushi {\em et~al.\/}(2018)Lushi, Goldstein \&
  Shelley]{lushi_nonlinear_2018}
{\sc \au{Lushi, E.}, \au{Goldstein, R.~E.} \& \au{Shelley, M.~J.}} \yr{2018}
  \at{Nonlinear concentration patterns and bands in autochemotactic
  suspensions}.  \jt{Phys. Rev. E}  \bvol{98}~(5),  \pg{052411}.

\bibitem[Lushi {\em et~al.\/}(2017)Lushi, Kantsler \&
  Goldstein]{lushi_scattering_2017}
{\sc \au{Lushi, E.}, \au{Kantsler, V.} \& \au{Goldstein, R.~E.}} \yr{2017}
  \at{Scattering of biflagellate microswimmers from surfaces}.  \jt{Phys. Rev.
  E}  \bvol{96}~(2),  \pg{023102}.

\bibitem[Lushi {\em et~al.\/}(2014)Lushi, Wioland \&
  Goldstein]{lushi_fluid_2014}
{\sc \au{Lushi, E.}, \au{Wioland, H.} \& \au{Goldstein, R.~E.}} \yr{2014}
  \at{Fluid flows created by swimming bacteria drive self-organization in
  confined suspensions}.  \jt{PNAS}  \bvol{111}~(27),  \pg{9733--9738}.

\bibitem[Makarchuk {\em et~al.\/}(2019)Makarchuk, Braz, Ara{\'u}jo, Ciric \&
  Volpe]{makarchuk_enhanced_2019}
{\sc \au{Makarchuk, S.}, \au{Braz, V.~C.}, \au{Ara{\'u}jo, N. A.~M.},
  \au{Ciric, L.} \& \au{Volpe, G.}} \yr{2019}  \at{Enhanced propagation of
  motile bacteria on surfaces due to forward scattering}.  \jt{Nat. Commun.}
  \bvol{10}~(1),  \pg{1--12}.

\bibitem[Makhnovskii(2019)]{makhnovskii_effect_2019}
{\sc \au{Makhnovskii, Y.~A.}} \yr{2019}  \at{Effect of particle size
  oscillations on drift and diffusion along a periodically corrugated channel}.
   \jt{Phys. Rev. E}  \bvol{99}~(3),  \pg{032102}.

\bibitem[Mathijssen {\em et~al.\/}(2019)Mathijssen, {Figueroa-Morales}, Junot,
  Cl{\'e}ment, Lindner \& Z{\"o}ttl]{mathijssen_oscillatory_2019}
{\sc \au{Mathijssen, A. J. T.~M.}, \au{{Figueroa-Morales}, N.}, \au{Junot, G.},
  \au{Cl{\'e}ment, {\'E}.}, \au{Lindner, A.} \& \au{Z{\"o}ttl, A.}} \yr{2019}
  \at{Oscillatory surface rheotaxis of swimming {{{\emph{E}}}}{\emph{. coli}}
  bacteria}.  \jt{Nat. Commun.}  \bvol{10}~(1),  \pg{1--12}.

\bibitem[Morris(2020)]{morris_shear_2020}
{\sc \au{Morris, J.~F.}} \yr{2020}  \at{Shear thickening of concentrated
  suspensions: recent developments and relation to other phenomena}.  \jt{Annu.
  Rev. Fluid Mech.}  \bvol{52}~(1),  \pg{121--144}.

\bibitem[Nambiar {\em et~al.\/}(2019)Nambiar, Phanikanth, Nott \&
  Subramanian]{nambiar_stress_2019}
{\sc \au{Nambiar, S.}, \au{Phanikanth, S.}, \au{Nott, P.~R.} \&
  \au{Subramanian, G.}} \yr{2019}  \at{Stress relaxation in a dilute bacterial
  suspension: the active\textendash passive transition}.  \jt{J. Fluid Mech.}
  \bvol{870},  \pg{1072--1104}.

\bibitem[Nili {\em et~al.\/}(2017)Nili, Kheyri, Abazari, Fahimniya \&
  Naji]{nili_population_2017}
{\sc \au{Nili, H.}, \au{Kheyri, M.}, \au{Abazari, J.}, \au{Fahimniya, A.} \&
  \au{Naji, A.}} \yr{2017}  \at{Population splitting of rodlike swimmers in
  {{Couette}} flow}.  \jt{Soft Matter}  \bvol{13}~(25),  \pg{4494--4506}.

\bibitem[Pedley \& Kessler(1990)]{pedley_new_1990}
{\sc \au{Pedley, T.~J.} \& \au{Kessler, J.~O.}} \yr{1990}  \at{A new continuum
  model for suspensions of gyrotactic micro-organisms}.  \jt{J. Fluid Mech.}
  \bvol{212},  \pg{155--182}.

\bibitem[Pedley \& Kessler(1992)]{pedley_hydrodynamic_1992}
{\sc \au{Pedley, T.~J.} \& \au{Kessler, J.~O.}} \yr{1992}  \at{Hydrodynamic
  phenomena in suspensions of swimming microorganisms}.  \jt{Annu. Rev. Fluid
  Mech.}  \bvol{24}~(1),  \pg{313--358}.

\bibitem[Peng \& Brady(2020)]{peng_upstream_2020}
{\sc \au{Peng, Z.} \& \au{Brady, J.~F.}} \yr{2020}  \at{Upstream swimming and
  {{Taylor}} dispersion of active {{Brownian}} particles}.  \jt{Phys. Rev.
  Fluids}  \bvol{5}~(7),  \pg{073102}.

\bibitem[Posten(2009)]{posten_design_2009}
{\sc \au{Posten, C.}} \yr{2009}  \at{Design principles of photo-bioreactors for
  cultivation of microalgae}.  \jt{Eng. Life Sci.}  \bvol{9}~(3),
  \pg{165--177}.

\bibitem[Romanczuk {\em et~al.\/}(2012)Romanczuk, B{\"a}r, Ebeling, Lindner \&
  {Schimansky-Geier}]{romanczuk_active_2012}
{\sc \au{Romanczuk, P.}, \au{B{\"a}r, M.}, \au{Ebeling, W.}, \au{Lindner, B.}
  \& \au{{Schimansky-Geier}, L.}} \yr{2012}  \at{Active {{Brownian}}
  particles}.  \jt{Eur. Phys. J. Spec. Top.}  \bvol{202}~(1),  \pg{1--162}.

\bibitem[{Rothschild}(1963)]{rothschild_non-random_1963}
{\sc \au{{Rothschild}}} \yr{1963}  \at{Non-random distribution of bull
  spermatozoa in a drop of sperm suspension}.  \jt{Nature}  \bvol{198}~(4886),
  \pg{1221--1222}.

\bibitem[Rusconi {\em et~al.\/}(2014)Rusconi, Guasto \&
  Stocker]{rusconi_bacterial_2014}
{\sc \au{Rusconi, R.}, \au{Guasto, J.~S.} \& \au{Stocker, R.}} \yr{2014}
  \at{Bacterial transport suppressed by fluid shear}.  \jt{Nat. Phys.}
  \bvol{10}~(3),  \pg{212--217}.

\bibitem[Saintillan(2018)]{saintillan_rheology_2018}
{\sc \au{Saintillan, D.}} \yr{2018}  \at{Rheology of active fluids}.  \jt{Annu.
  Rev. Fluid Mech.}  \bvol{50}~(1),  \pg{563--592}.

\bibitem[Saintillan \& Shelley(2013)]{saintillan_active_2013}
{\sc \au{Saintillan, D.} \& \au{Shelley, M.~J.}} \yr{2013}  \at{Active
  suspensions and their nonlinear models}.  \jt{Comptes Rendus Phys.}
  \bvol{14}~(6),  \pg{497--517}.

\bibitem[Sandoval \& Dagdug(2014)]{sandoval_effective_2014}
{\sc \au{Sandoval, M.} \& \au{Dagdug, L.}} \yr{2014}  \at{Effective diffusion
  of confined active {{Brownian}} swimmers}.  \jt{Phys. Rev. E}  \bvol{90}~(6),
   \pg{062711}.

\bibitem[Sandoval {\em et~al.\/}(2014)Sandoval, Marath, Subramanian \&
  Lauga]{sandoval_stochastic_2014}
{\sc \au{Sandoval, M.}, \au{Marath, N.~K.}, \au{Subramanian, G.} \& \au{Lauga,
  E.}} \yr{2014}  \at{Stochastic dynamics of active swimmers in linear flows}.
  \jt{J. Fluid Mech.}  \bvol{742},  \pg{50--70}.

\bibitem[Schweitzer(2003)]{schweitzer_brownian_2003}
{\sc \au{Schweitzer, F.}} \yr{2003} {\em Brownian agents and active particles:
  collective dynamics in the natural and social sciences\/}.  \publ{{Berlin}:
  {Springer}}.

\bibitem[Sipos {\em et~al.\/}(2015)Sipos, Nagy, Di~Leonardo \&
  Galajda]{sipos_hydrodynamic_2015}
{\sc \au{Sipos, O.}, \au{Nagy, K.}, \au{Di~Leonardo, R.} \& \au{Galajda, P.}}
  \yr{2015}  \at{Hydrodynamic trapping of swimming bacteria by convex walls}.
  \jt{Phys. Rev. Lett.}  \bvol{114}~(25),  \pg{258104}.

\bibitem[Spagnolie \& Lauga(2012)]{spagnolie_hydrodynamics_2012}
{\sc \au{Spagnolie, S.~E.} \& \au{Lauga, E.}} \yr{2012}  \at{Hydrodynamics of
  self-propulsion near a boundary: predictions and accuracy of far-field
  approximations}.  \jt{J. Fluid Mech.}  \bvol{700},  \pg{105--147}.

\bibitem[Strand {\em et~al.\/}(1987)Strand, Kim \&
  Karrila]{strand_computation_1987}
{\sc \au{Strand, S.~R.}, \au{Kim, S.} \& \au{Karrila, S.~J.}} \yr{1987}
  \at{Computation of rheological properties of suspensions of rigid rods:
  stress growth after inception of steady shear flow}.  \jt{J. Non-Newton.
  Fluid Mech.}  \bvol{24}~(3),  \pg{311--329}.

\bibitem[Taghizadeh {\em et~al.\/}(2020)Taghizadeh, {Vald{\'e}s-Parada} \&
  Wood]{taghizadeh_preasymptotic_2020}
{\sc \au{Taghizadeh, E.}, \au{{Vald{\'e}s-Parada}, F.~J.} \& \au{Wood, B.~D.}}
  \yr{2020}  \at{Preasymptotic {{Taylor}} dispersion: evolution from the
  initial condition}.  \jt{J. Fluid Mech.}  \bvol{889},  \pg{A5}.

\bibitem[Takatori \& Brady(2017)]{takatori_superfluid_2017}
{\sc \au{Takatori, S.~C.} \& \au{Brady, J.~F.}} \yr{2017}  \at{Superfluid
  behavior of active suspensions from diffusive stretching}.  \jt{Phys. Rev.
  Lett.}  \bvol{118}~(1),  \pg{018003}.

\bibitem[Taylor(1953)]{taylor_dispersion_1953}
{\sc \au{Taylor, G.}} \yr{1953}  \at{Dispersion of soluble matter in solvent
  flowing slowly through a tube}.  \jt{Proc. R. Soc. Lond. Math. Phys. Eng.
  Sci.}  \bvol{219}~(1137),  \pg{186--203}.

\bibitem[Taylor(1954)]{taylor_dispersion_1954}
{\sc \au{Taylor, G.}} \yr{1954}  \at{The dispersion of matter in turbulent flow
  through a pipe}.  \jt{Proc. R. Soc. Lond. Math. Phys. Eng. Sci.}
  \bvol{223}~(1155),  \pg{446--468}.

\bibitem[{ten Hagen} {\em et~al.\/}(2011{\natexlab{{\em a\/}}}){ten Hagen},
  {van Teeffelen} \& L{\"o}wen]{tenhagen_brownian_2011}
{\sc \au{{ten Hagen}, B.}, \au{{van Teeffelen}, S.} \& \au{L{\"o}wen, H.}}
  \yr{2011{\natexlab{{\em a\/}}}}  \at{Brownian motion of a self-propelled
  particle}.  \jt{J. Phys. Condens. Matter}  \bvol{23}~(19),  \pg{194119}.

\bibitem[{ten Hagen} {\em et~al.\/}(2011{\natexlab{{\em b\/}}}){ten Hagen},
  Wittkowski \& L{\"o}wen]{tenhagen_brownian_2011a}
{\sc \au{{ten Hagen}, B.}, \au{Wittkowski, R.} \& \au{L{\"o}wen, H.}}
  \yr{2011{\natexlab{{\em b\/}}}}  \at{Brownian dynamics of a self-propelled
  particle in shear flow}.  \jt{Phys. Rev. E}  \bvol{84}~(3),  \pg{031105}.

\bibitem[Uspal {\em et~al.\/}(2015)Uspal, Popescu, Dietrich \&
  Tasinkevych]{uspal_rheotaxis_2015}
{\sc \au{Uspal, W.~E.}, \au{Popescu, M.~N.}, \au{Dietrich, S.} \&
  \au{Tasinkevych, M.}} \yr{2015}  \at{Rheotaxis of spherical active particles
  near a planar wall}.  \jt{Soft Matter}  \bvol{11}~(33),  \pg{6613--6632}.

\bibitem[Vedel {\em et~al.\/}(2014)Vedel, Hovad \&
  Bruus]{vedel_time-dependent_2014}
{\sc \au{Vedel, S.}, \au{Hovad, E.} \& \au{Bruus, H.}} \yr{2014}
  \at{Time-dependent {{Taylor}}\textendash{{Aris}} dispersion of an initial
  point concentration}.  \jt{J. Fluid Mech.}  \bvol{752},  \pg{107--122}.

\bibitem[Vennamneni {\em et~al.\/}(2020)Vennamneni, Nambiar \&
  Subramanian]{vennamneni_shear-induced_2020}
{\sc \au{Vennamneni, L.}, \au{Nambiar, S.} \& \au{Subramanian, G.}} \yr{2020}
  \at{Shear-induced migration of microswimmers in pressure-driven channel
  flow}.  \jt{J. Fluid Mech.}  \bvol{890},  \pg{A15}.

\bibitem[Vicsek \& Zafeiris(2012)]{vicsek_collective_2012}
{\sc \au{Vicsek, T.} \& \au{Zafeiris, A.}} \yr{2012}  \at{Collective motion}.
  \jt{Phys. Rep.}  \bvol{517}~(3),  \pg{71--140}.

\bibitem[Volpe {\em et~al.\/}(2011)Volpe, Buttinoni, Vogt, K{\"u}mmerer \&
  Bechinger]{volpe_microswimmers_2011}
{\sc \au{Volpe, G.}, \au{Buttinoni, I.}, \au{Vogt, D.}, \au{K{\"u}mmerer,
  H.-J.} \& \au{Bechinger, C.}} \yr{2011}  \at{Microswimmers in patterned
  environments}.  \jt{Soft Matter}  \bvol{7}~(19),  \pg{8810--8815}.

\bibitem[Volpe {\em et~al.\/}(2014)Volpe, Gigan \&
  Volpe]{volpe_simulation_2014}
{\sc \au{Volpe, G.}, \au{Gigan, S.} \& \au{Volpe, G.}} \yr{2014}
  \at{Simulation of the active {{Brownian}} motion of a microswimmer}.  \jt{Am.
  J. Phys.}  \bvol{82}~(7),  \pg{659--664}.

\bibitem[Wang {\em et~al.\/}(2020)Wang, Jiang, Chen, Li \&
  Tao]{wang_vertical_2020}
{\sc \au{Wang, B.}, \au{Jiang, W.}, \au{Chen, G.}, \au{Li, Z.} \& \au{Tao, L.}}
  \yr{2020}  \at{Vertical distribution and longitudinal dispersion of
  gyrotactic microorganisms in an open channel flow}.  \jt{Phys. Rev. Fluids}
  ~(submitted).

\bibitem[Wang \& Chen(2017)]{wang_basic_2017}
{\sc \au{Wang, P.} \& \au{Chen, G.~Q.}} \yr{2017}  \at{Basic characteristics of
  {{Taylor}} dispersion in a laminar tube flow with wall absorption: exchange
  rate, advection velocity, dispersivity, skewness and kurtosis in their full
  time dependance}.  \jt{Int. J. Heat Mass Transf.}  \bvol{109},
  \pg{844--852}.

\bibitem[Wu \& Chen(2014)]{wu_approach_2014}
{\sc \au{Wu, Z.} \& \au{Chen, G.~Q.}} \yr{2014}  \at{Approach to transverse
  uniformity of concentration distribution of a solute in a solvent flowing
  along a straight pipe}.  \jt{J. Fluid Mech.}  \bvol{740},  \pg{196--213}.

\bibitem[Xiao {\em et~al.\/}(2019)Xiao, Wei \& Wang]{xiao_review_2019}
{\sc \au{Xiao, Z.}, \au{Wei, M.} \& \au{Wang, W.}} \yr{2019}  \at{A review of
  micromotors in confinements: pores, channels, grooves, steps, interfaces,
  chains, and swimming in the bulk}.  \jt{ACS Appl. Mater. Interfaces}
  \bvol{11}~(7),  \pg{6667--6684}.

\bibitem[Yariv \& Schnitzer(2014)]{yariv_ratcheting_2014}
{\sc \au{Yariv, E.} \& \au{Schnitzer, O.}} \yr{2014}  \at{Ratcheting of
  {{Brownian}} swimmers in periodically corrugated channels: a reduced
  {{Fokker}}-{{Planck}} approach}.  \jt{Phys. Rev. E}  \bvol{90}~(3),
  \pg{032115}.

\bibitem[Yasa {\em et~al.\/}(2018)Yasa, Erkoc, Alapan \&
  Sitti]{yasa_microalga-powered_2018}
{\sc \au{Yasa, O.}, \au{Erkoc, P.}, \au{Alapan, Y.} \& \au{Sitti, M.}}
  \yr{2018}  \at{Microalga-powered microswimmers toward active cargo delivery}.
   \jt{Adv. Mater.}  \bvol{30}~(45),  \pg{1804130}.

\bibitem[Zeng \& Pedley(2018)]{zeng_distribution_2018}
{\sc \au{Zeng, L.} \& \au{Pedley, T.~J.}} \yr{2018}  \at{Distribution of
  gyrotactic micro-organisms in complex three-dimensional flows. {{Part}} 1.
  {{Horizontal}} shear flow past a vertical circular cylinder}.  \jt{J. Fluid
  Mech.}  \bvol{852},  \pg{358--397}.

\bibitem[Zheng {\em et~al.\/}(2013)Zheng, {ten Hagen}, Kaiser, Wu, Cui,
  {Silber-Li} \& L{\"o}wen]{zheng_non-gaussian_2013}
{\sc \au{Zheng, X.}, \au{{ten Hagen}, B.}, \au{Kaiser, A.}, \au{Wu, M.},
  \au{Cui, H.}, \au{{Silber-Li}, Z.} \& \au{L{\"o}wen, H.}} \yr{2013}
  \at{Non-{{Gaussian}} statistics for the motion of self-propelled {{Janus}}
  particles: experiment versus theory}.  \jt{Phys. Rev. E}  \bvol{88}~(3),
  \pg{032304}.

\bibitem[Z{\"o}ttl \& Stark(2012)]{zottl_nonlinear_2012}
{\sc \au{Z{\"o}ttl, A.} \& \au{Stark, H.}} \yr{2012}  \at{Nonlinear dynamics of
  a microswimmer in {{Poiseuille}} flow}.  \jt{Phys. Rev. Lett.}
  \bvol{108}~(21),  \pg{218104}.

\bibitem[Z{\"o}ttl \& Stark(2013)]{zottl_periodic_2013}
{\sc \au{Z{\"o}ttl, A.} \& \au{Stark, H.}} \yr{2013}  \at{Periodic and
  quasiperiodic motion of an elongated microswimmer in {{Poiseuille}} flow}.
  \jt{Eur. Phys. J. E}  \bvol{36}~(1),  \pg{4}.

\end{thebibliography}
%\bibliography{mylibraryAbbreviatedExp}

\end{document}